\documentclass{article}
\usepackage{amsmath}
\usepackage{graphicx}
\usepackage{subfigure}

\title{A Mathematical Framework Exhibiting the Emergence of Dynamic Expansion of Task Repertoire in \emph{Pheidole dentata}}
\author{Jason M.\ Graham$^{1,3,\ast}$, Ivan L.\ Simpson-Kent$^{3}$, Marc A.\ Seid$^{2,3}$ }
\date{ }
\begin{document}
\maketitle

\noindent ${ }^{1}$ Department of Mathematics, University of Scranton, Scranton,  PA, 18510 USA \\
\noindent ${ }^{2}$ Department of Biology, University of Scranton, Scranton,  PA, 18510 USA \\
\noindent ${ }^{3}$ Neuroscience Program, University of Scranton, Scranton,  PA, 18510 USA \\
\noindent $\ast$ Corresponding Author E-mail: jason.graham@scranton.edu

\begin{abstract}
  The division of labor (DOL) and task allocation among groups of ants living in a colony is thought to be highly efficient, and key to the robust survival of a colony. A great deal of experimental and theoretical work has been done toward gaining a clear understanding of the evolution of, and underlying mechanisms of these phenomena.  Much of this research has utilized mathematical modeling. Here we continue this tradition by developing a mathematical model for a particular aspect of task allocation, known as age-related repertoire expansion, that has been observed in the minor workers of the ant species \emph{Pheidole dentata}. In fact, we present a relatively broad mathematical modeling framework based on the dynamics of the frequency with which members of specific age groups carry out distinct tasks. We apply our modeling approach to a specific task allocation scenario, and compare our theoretical results with experimental data. It is observed that the model predicts perceived behavior, and provides a possible explanation for the aforementioned experimental results.
\end{abstract}

{\bf Keywords:}  Repertoire expansion; Task allocation; Division of labor; Social insects; Temporal polyethism; Dynamical systems

\section{Introduction}
\label{intro}

Social insects are renown for their complex and highly organized behaviors \cite{holldobler2009}. This is particularly true of ants, and there is an extensive research literature detailing investigations of many aspects of the individual and collective behaviors observed of ants living in a colony, see \emph{e.g.}\ \cite{bourke1995,garnier2007,holldobler1990,holldobler2009,oster1978,sudd2013}. One thing that is apparent is that the collective behavior of ants has played a principal role in the evolutionary success of this group. This is especially the case regarding the phenomena of division of labor (DOL) and task allocation (TA) among workers in a colony \cite{beshers2001,bonabeau1999,bourke1995,cornejo2014,duarte2011,franks1994,gordon1989,oster1978,sudd2013,sumpter2010}.

The seemingly highly organized and efficient task allocation and division of labor makes colonies robust with respect to their ability to collectively respond to survival needs. Many ideas have been developed to explain the behavioral and physiological mechanisms that underlie social insect work dynamics. Among these are the theory of castes,\emph{e.g.}\ \cite{oster1978,sudd2013}; the idea of foraging for work, \emph{e.g.}\ \cite{franks1994}; and the idea of a fixed threshold response mechanism, \emph{e.g.}\ \cite{bonabeau1998}. The review of Beshers and Fewell describes most of the prominent theoretical models for ant task allocation \cite{beshers2001}. Along with many of these ideas are corresponding mathematical models that serve to aid in formulating and testing the theoretical ideas \cite{assis2009,beshers2001,bonabeau1996,bonabeau1998,cornejo2014,gordon1999,oster1978,sumpter2010,udiani2015}.

One thing is clear, in many species of ant, age plays some significant role in the allocation of specific tasks among the worker class, a feature known as temporal polyethism, \cite{holldobler1990,holldobler2009,oster1978,sudd2013}. This is, in particular, the case with the widely studied species \emph{Pheidole dentata}. For example, normally, foraging is carried out by the oldest workers, while nursing and other forms of brood-care are carried out by the youngest workers. Furthermore, in many ant species, the number or types of primary tasks a worker performs evolves with age. However, many of the details that underlie temporal polyethism and age-based worker castes remain to be clarified. For example, the degree of task overlap among age-groups is unclear.

There are two classical extreme cases of age-based task allocation that were carefully described by Wilson and others, see \emph{e.g.}\ \cite{holldobler1990,holldobler2009,oster1978,sudd2013}. These are the discrete case, in which each of several distinct sets of tasks are matched one-to-one with several distinct age groups. Then, there is the continuous case, in which there is a smooth, overlapping transition in the sets of tasks that are performed across increasing ages\footnote{Clear illustrations of each case are exhibited bellow. See also the discussion in \cite{oster1978}.}.

More recently, a phenomenon known as repertoire expansion has been observed in experiments with the ant species \emph{Pheidole dentata} \cite{brown1998,calabi1989,calabi1983,seid2008,seid2006}. This may be viewed as a specific instance of the continuous case of temporal polyethism. What is peculiar to repertoire expansion, is that ageing workers do not cease to perform tasks, such as brood-care, but simply expand the repertoire of tasks that they perform with age \cite{seid2008,seid2006}. In other words, as minor workers of \emph{Pheidole dentata} age, they take on new tasks, while simultaneously retaining a memory for tasks they previously performed, and often resort to carrying out these former tasks as the need arises. Thus, there is an intrinsic plasticity exhibited by older workers, while younger workers tend to maintain a greater degree of specialization.

\begin{figure}[h!]
             \centering
             \subfigure{\includegraphics[width=1.15in,height=1.15in]{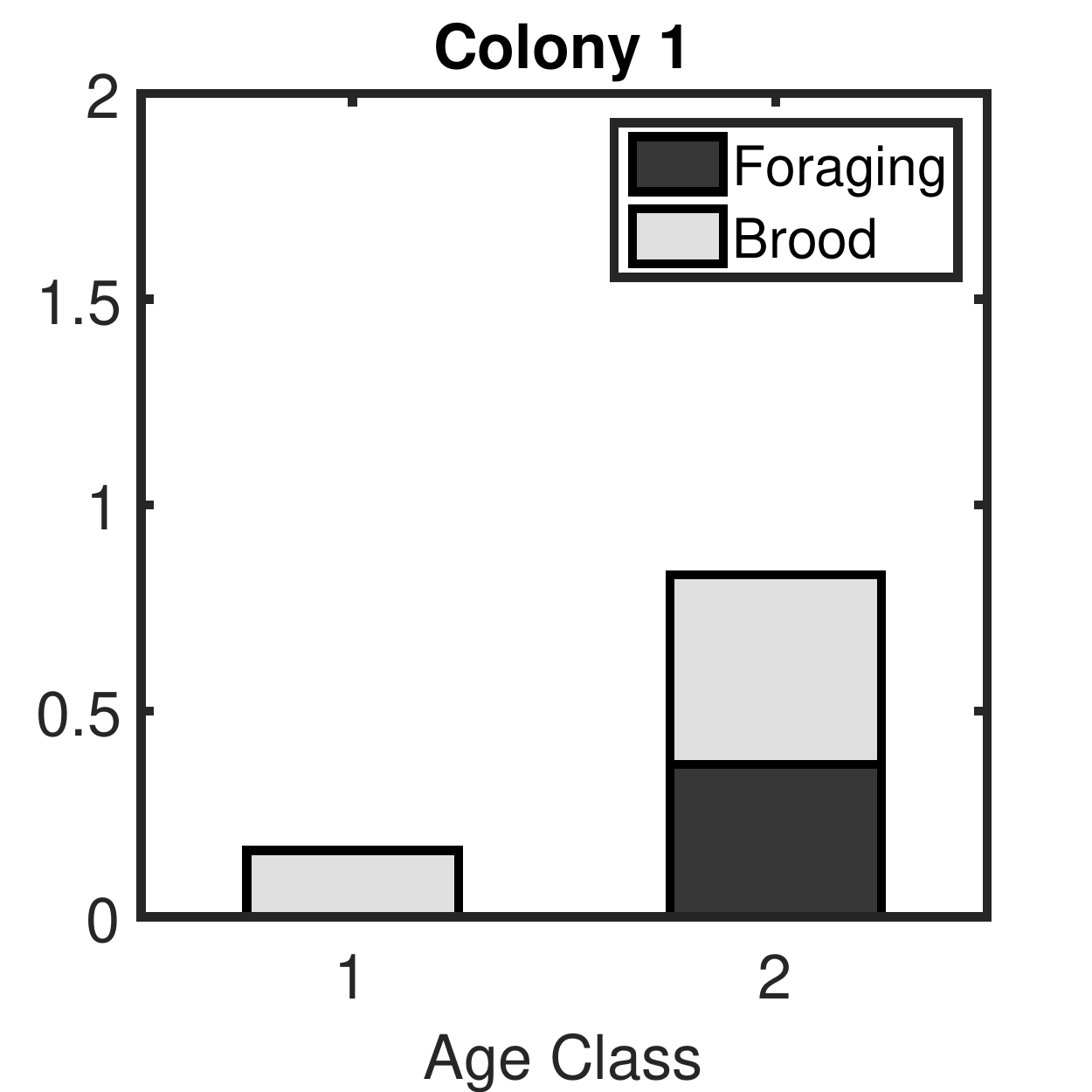}}
             \subfigure{\includegraphics[width=1.15in,height=1.15in]{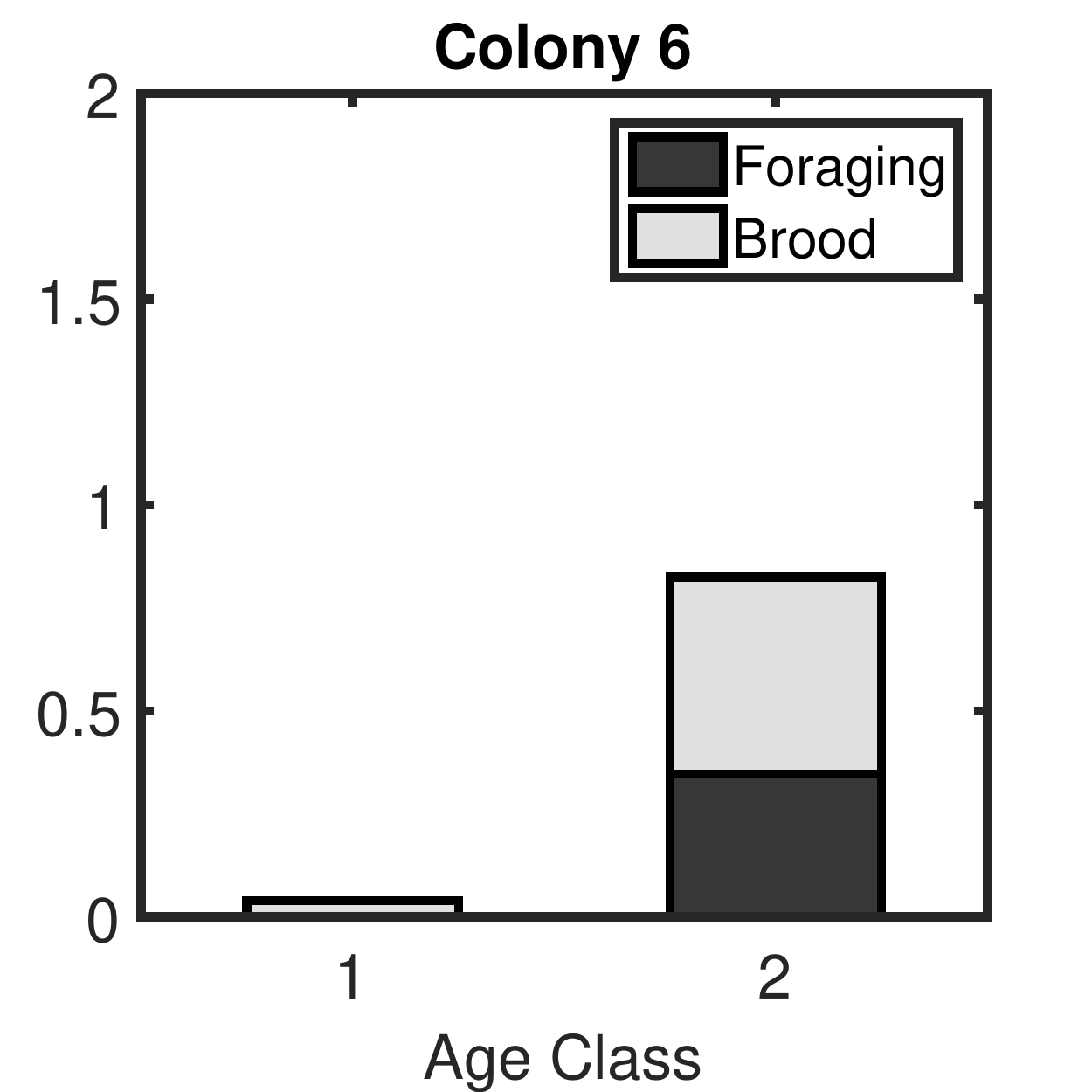}} \\
             \subfigure{\includegraphics[width=1.15in,height=1.15in]{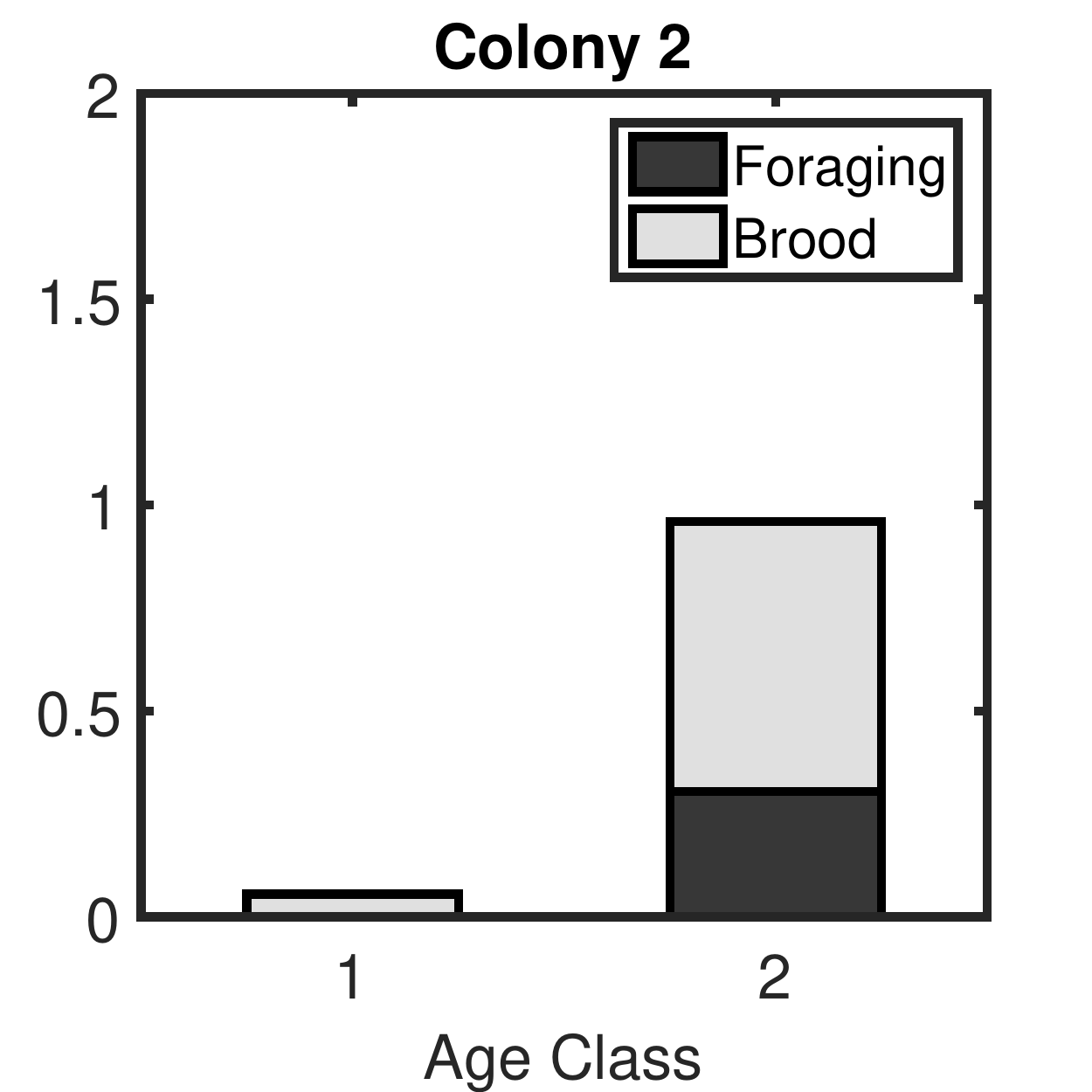}}
             \subfigure{\includegraphics[width=1.15in,height=1.15in]{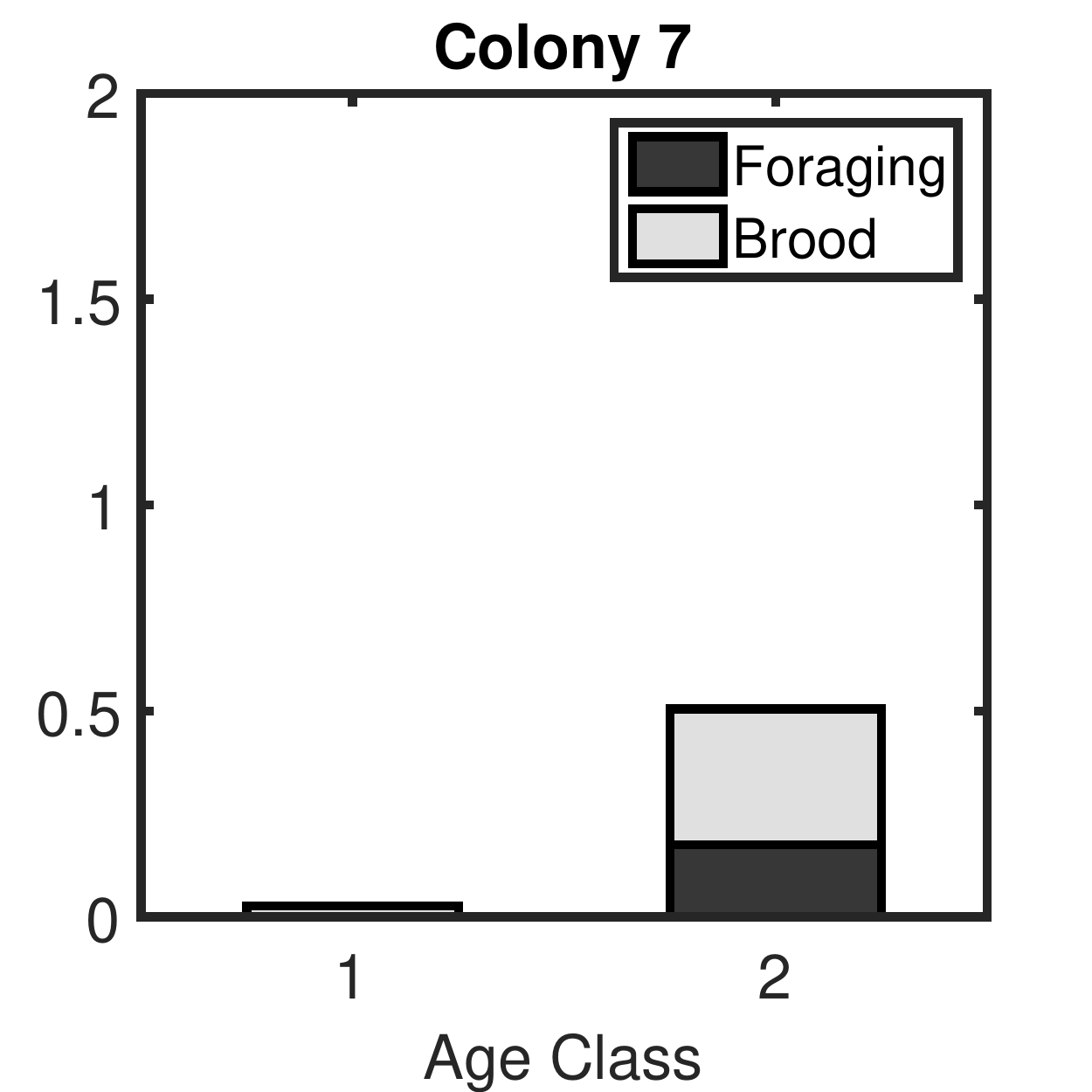}} \\
             \subfigure{\includegraphics[width=1.15in,height=1.15in]{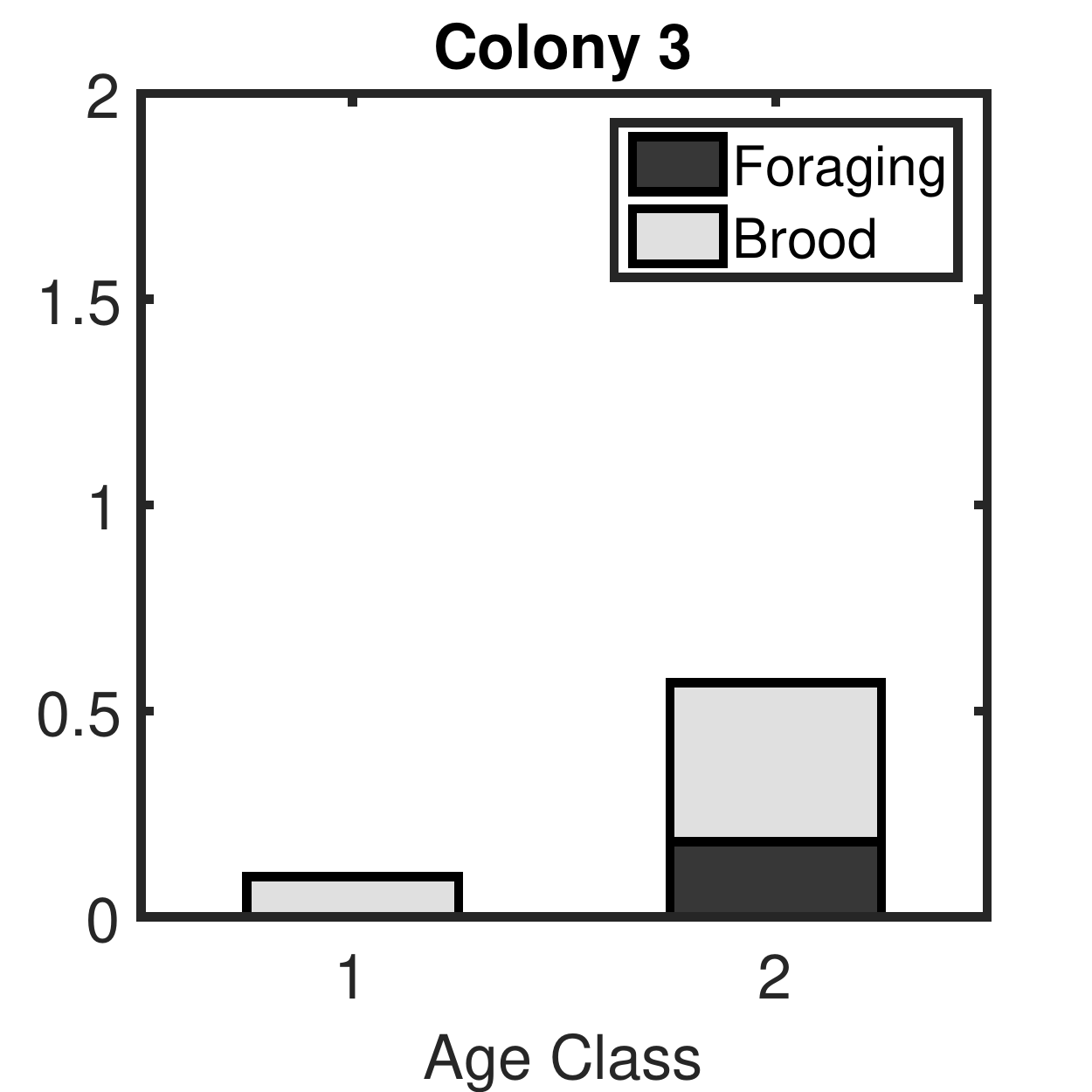}}
             \subfigure{\includegraphics[width=1.15in,height=1.15in]{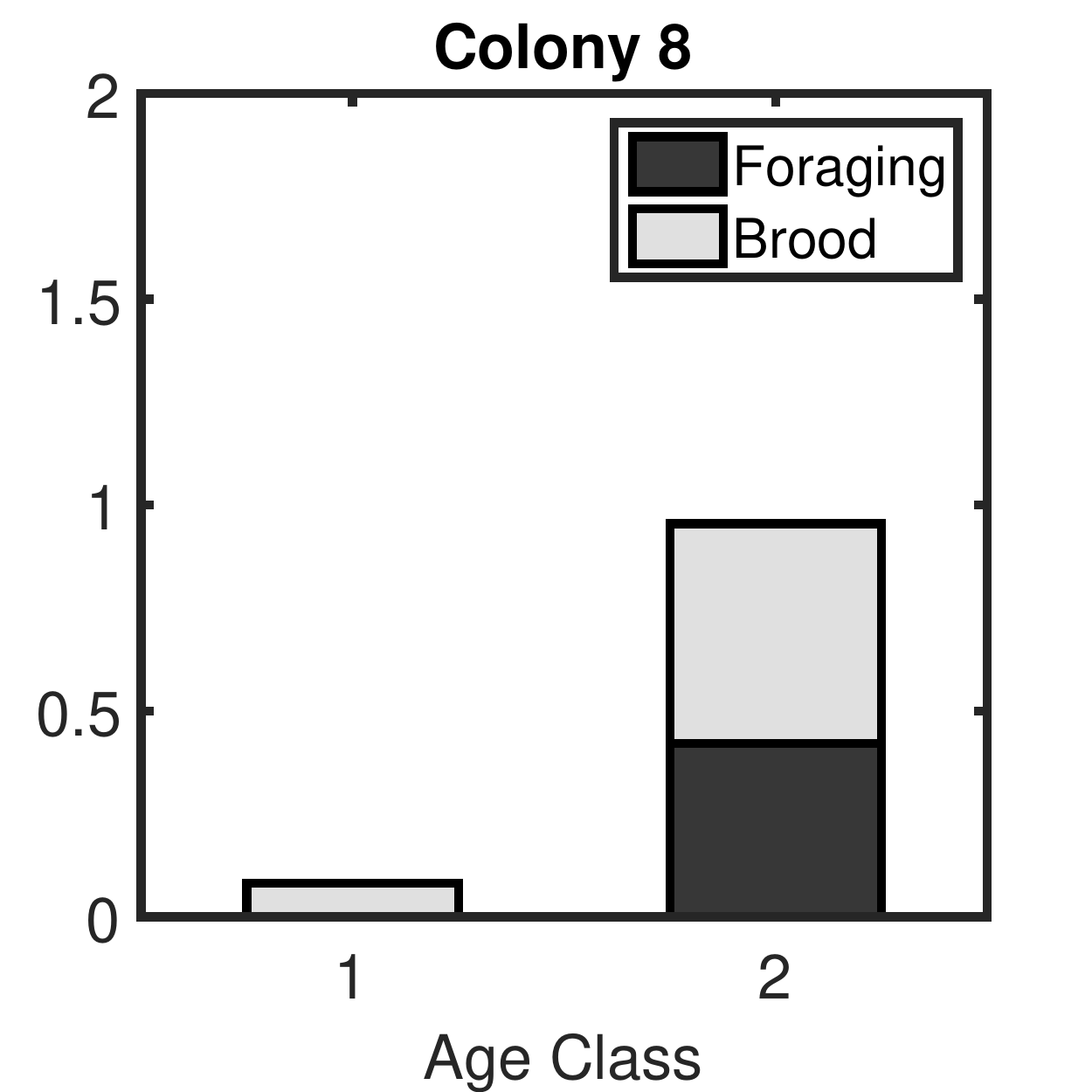}} \\
             \subfigure{\includegraphics[width=1.15in,height=1.15in]{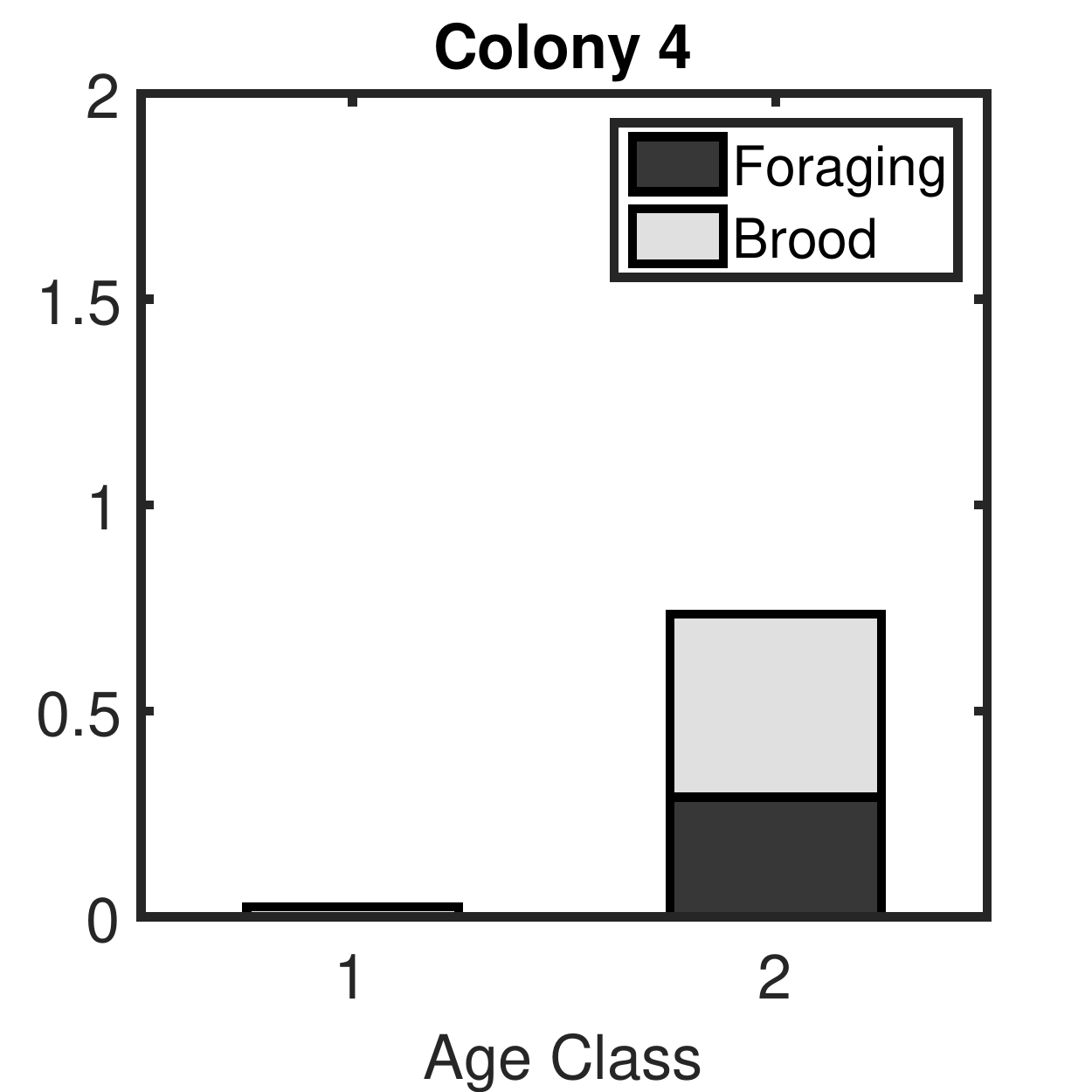}}
             \subfigure{\includegraphics[width=1.15in,height=1.15in]{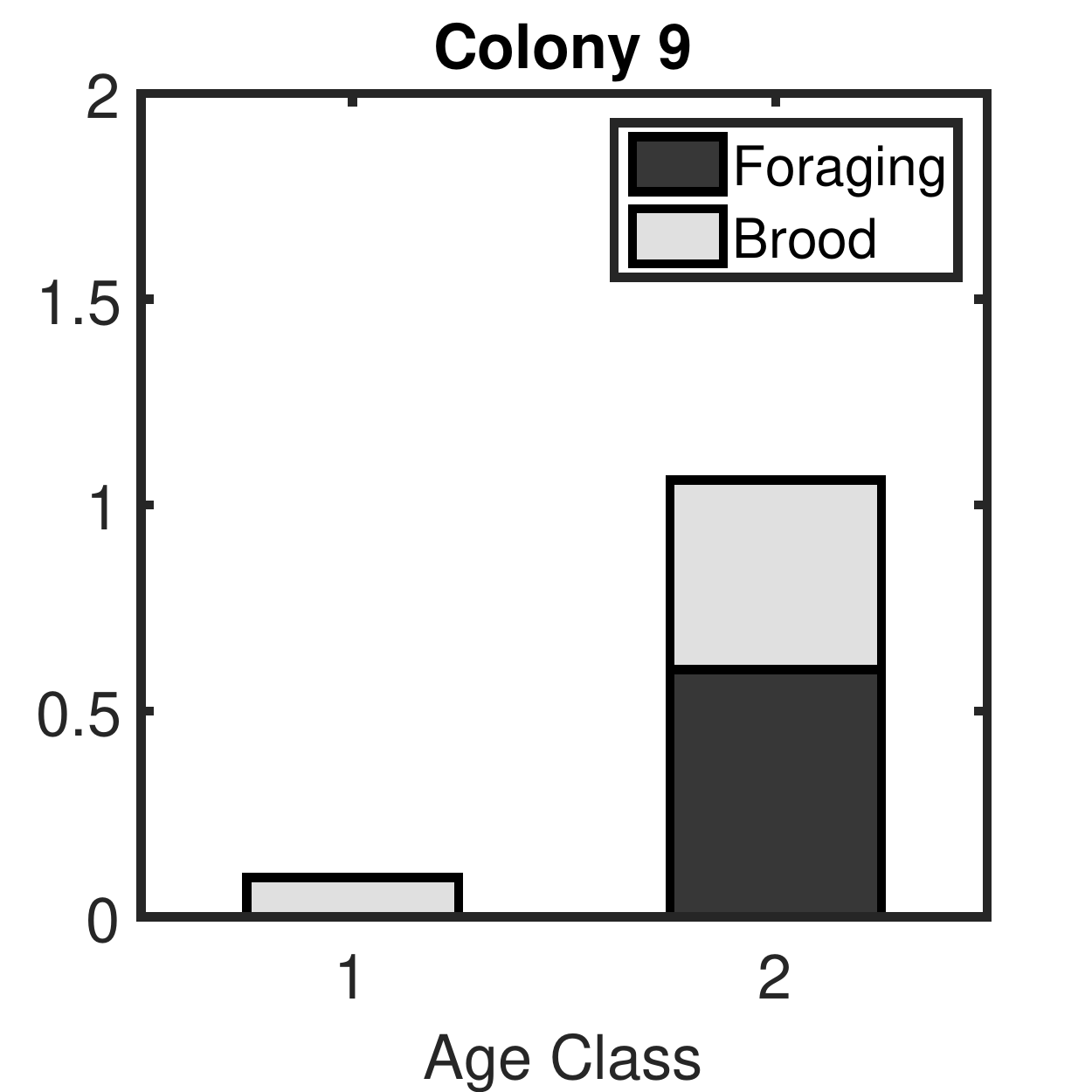}} \\
             \subfigure{\includegraphics[width=1.15in,height=1.15in]{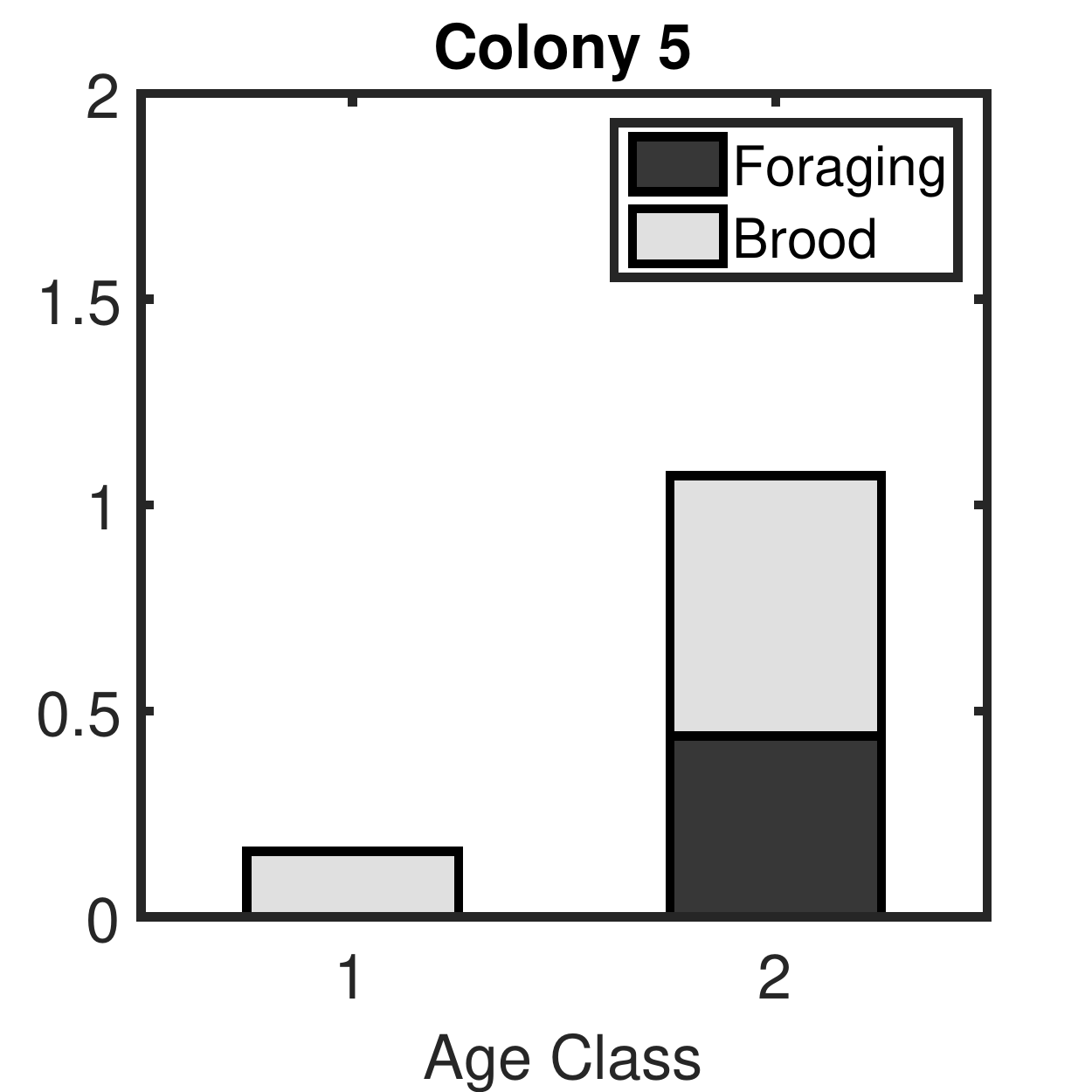}}
             \subfigure{\includegraphics[width=1.15in,height=1.15in]{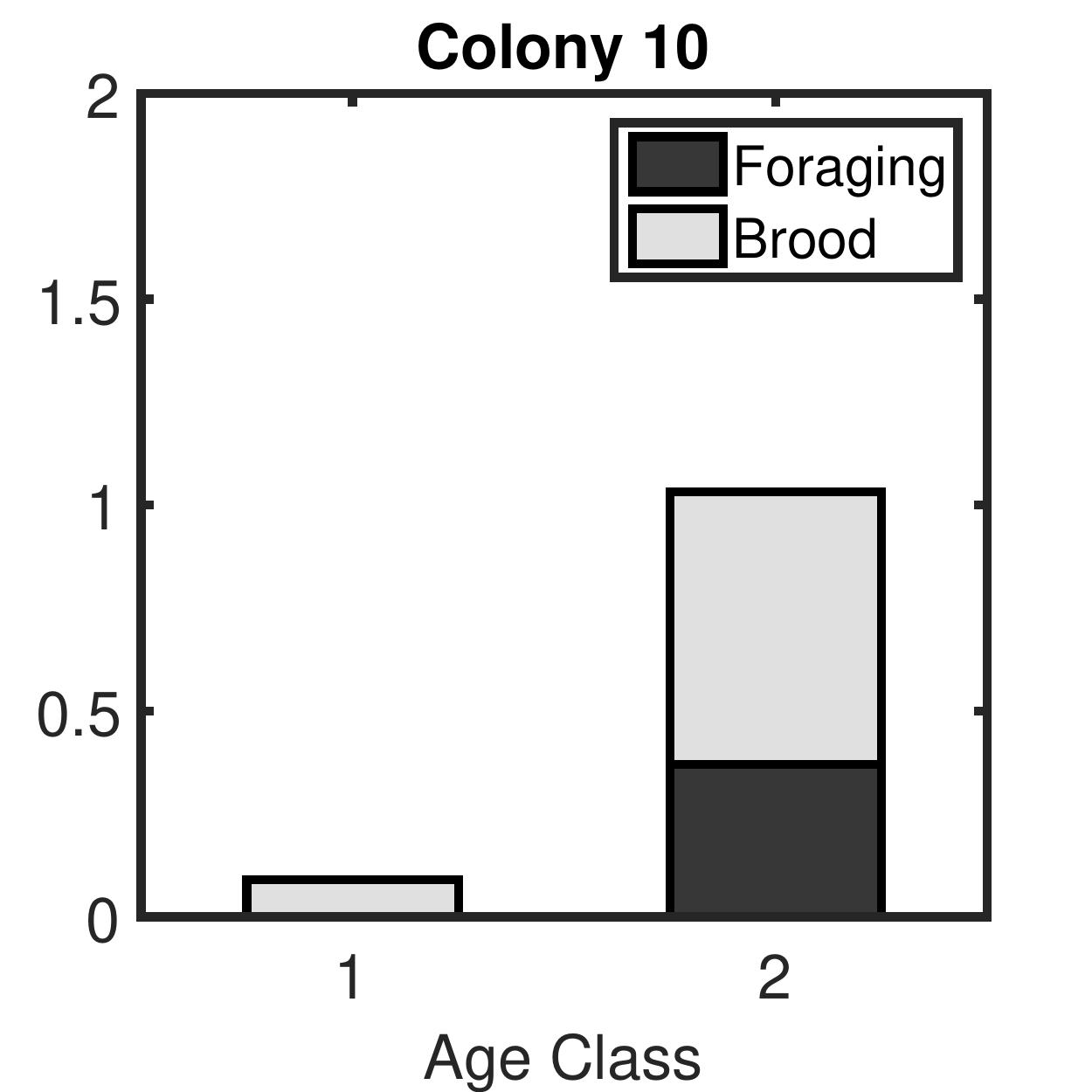}}
             \caption{Data, from \cite{seid2006}, combined and rescaled to have two tasks and two age groups. This exhibits repertoire expansion, or task stacking for two age-groups that perform two distinct task, but with one age-group highly specialized to a specific task. This is the actual data from \cite{seid2006} but with focus only on broad task classification in terms of brood-care and foraging.}\label{expansion}
  \end{figure}

Consider, as an exemplary case, the tasks of brood-care and foraging as carried out by the workers of the species \emph{Pheidole dentata}. When one carefully tracks the frequencies with which two distinct age-classes, a young ($<$ 20 days) and an old ($>$ 20 days), carry out these respective tasks, the data shows a phenomenon of ``task-stacking'' which is clearly exhibited in figure \ref{expansion}. This is what we mean by repertoire expansion. The data used to produce figure \ref{expansion} is that of \cite{seid2006}, but presented with respect to a broader task classification. When greater detail is taken into consideration, as it is in \cite{seid2006}, task-stacking and repertoire expansion is even more pronounced. It is the dynamics that underlie this phenomenon that we seek to represent.

The remainder of the paper proceeds as follows: In the next section we describe, generally, our approach in developing a mathematical model that allows for a representation of repertoire expansion. Section \ref{firstApplication} provides an application of a special case of the general framework presented in section \ref{methods}. There, we also make connections with the experimental results presented in \cite{seid2006}. Finally, the paper concludes with a discussion of our observations, connections with other works, and future directions and open problems.

\section{Methods}
\label{methods}

To our knowledge, there is currently no existing mathematical model for ant DOL or TA that can be used to directly represent the repertoire expansion data from \cite{seid2008,seid2006}. It is the primary goal of this work to develop such a mathematical modeling approach. Furthermore, we attempt to do so in a broad manner, so that other observed DOL related features may arise as emergent phenomena, at least under certain parameter settings. We note that the very recent mathematical work in \cite{kang2015} may provide significant insight for age-polyethism.

Specifically, we introduce the {\bf age-task frequency matrix}
\begin{equation} \Omega = (\omega_{ij}) = \left(\begin{array}{cccc} \omega_{11} & \omega_{12} & \cdots & \omega_{1T} \\ \omega_{21} & \omega_{22} & \cdots & \omega_{2T} \\ \vdots & \ddots & \cdots & \vdots \\ \omega_{A1} & \omega_{A2} & \cdots & \omega_{AT}  \end{array}  \right) ,\label{eq:ATmat}\end{equation}
where the $i,j$-th entry, $\omega_{ij}$, represents the frequency with which members of age group $i$ perform task $j$. Here $A$ denotes the total number of distinct age-groups, while $T$ denotes the total number of distinct tasks. We note that, in general, each entry $\omega_{ij}$, and hence the entire matrix $\Omega$ is expected to depend on some number of independent variables such as time, colony size, population, etc.; and may even be coupled with population models for instance, should the need arise. In the sequel, we establish a system of ordinary differential equations (ODEs) that allow for the calculation of specific values for the $\omega_{ij}$ as a function of a single independent variable which we interpret to be time. Throughout, we work with scaled values so that it is always the case that $0\leq \omega_{ij} \leq 1$.

Our motivation for taking the age-task frequency as our basic theoretical quantity is manifold. Firstly, we feel that it is closely connected with the way in which the experimental observations of \cite{seid2006} that we seek to model are quantified, see also \cite{seeley1982}. Furthermore, considering frequency, as we do, allows for a more simple accounting of efficiency than when compared with tracking individuals, which is difficult to do in many experimental setups with a large colony or population.

In the remainder of this section, we establish our approach to constructing equations that allow for the exhibition of the dynamics for the $\omega_{ij}$. But first, we briefly describe some of the utilities of the approach we have developed.

\begin{figure}[h!]
   \centering
   \subfigure[ ]{\includegraphics[height=1.25in]{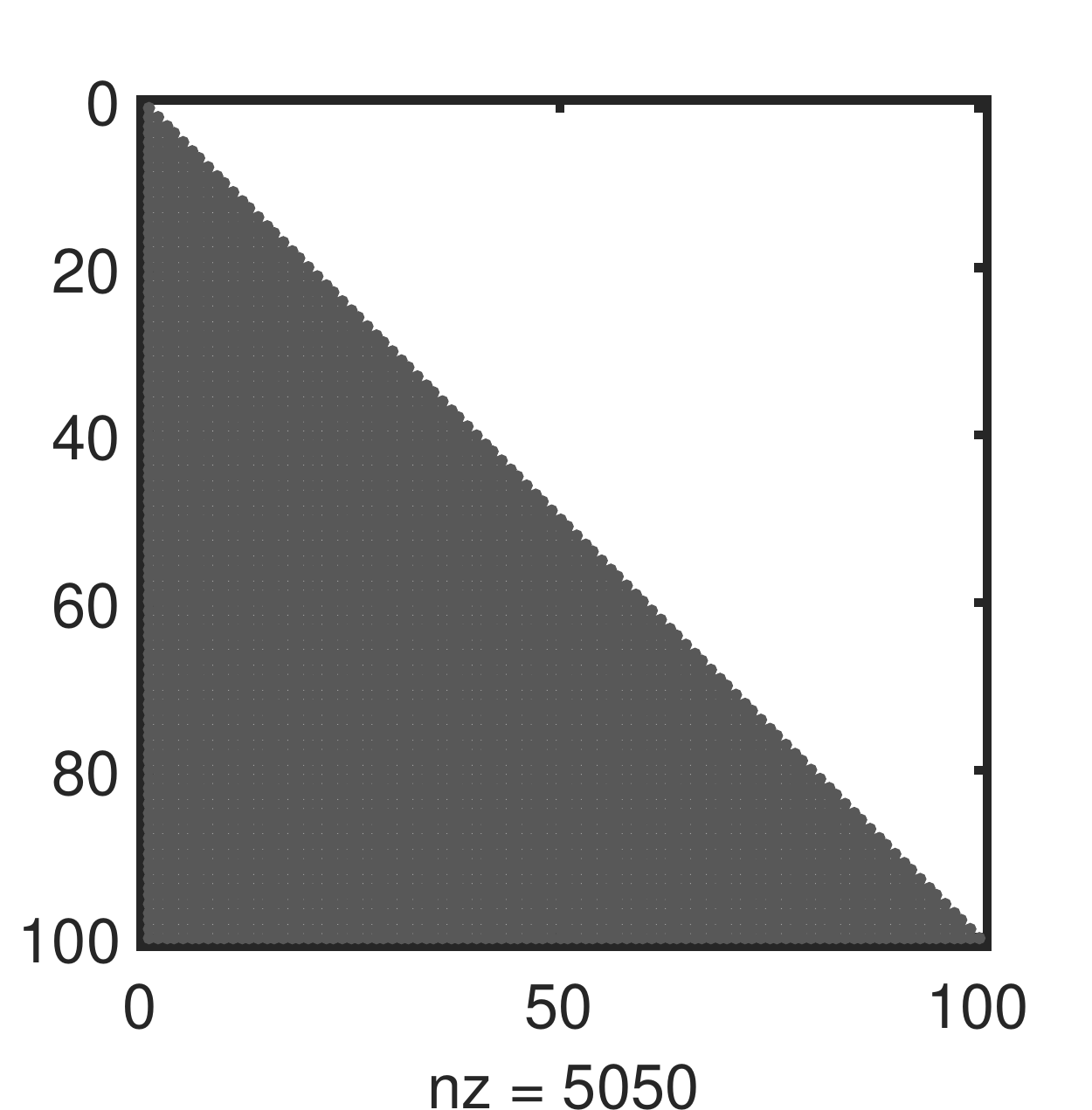}\label{figOneA}}
   \subfigure[ ]{\includegraphics[height=1.25in]{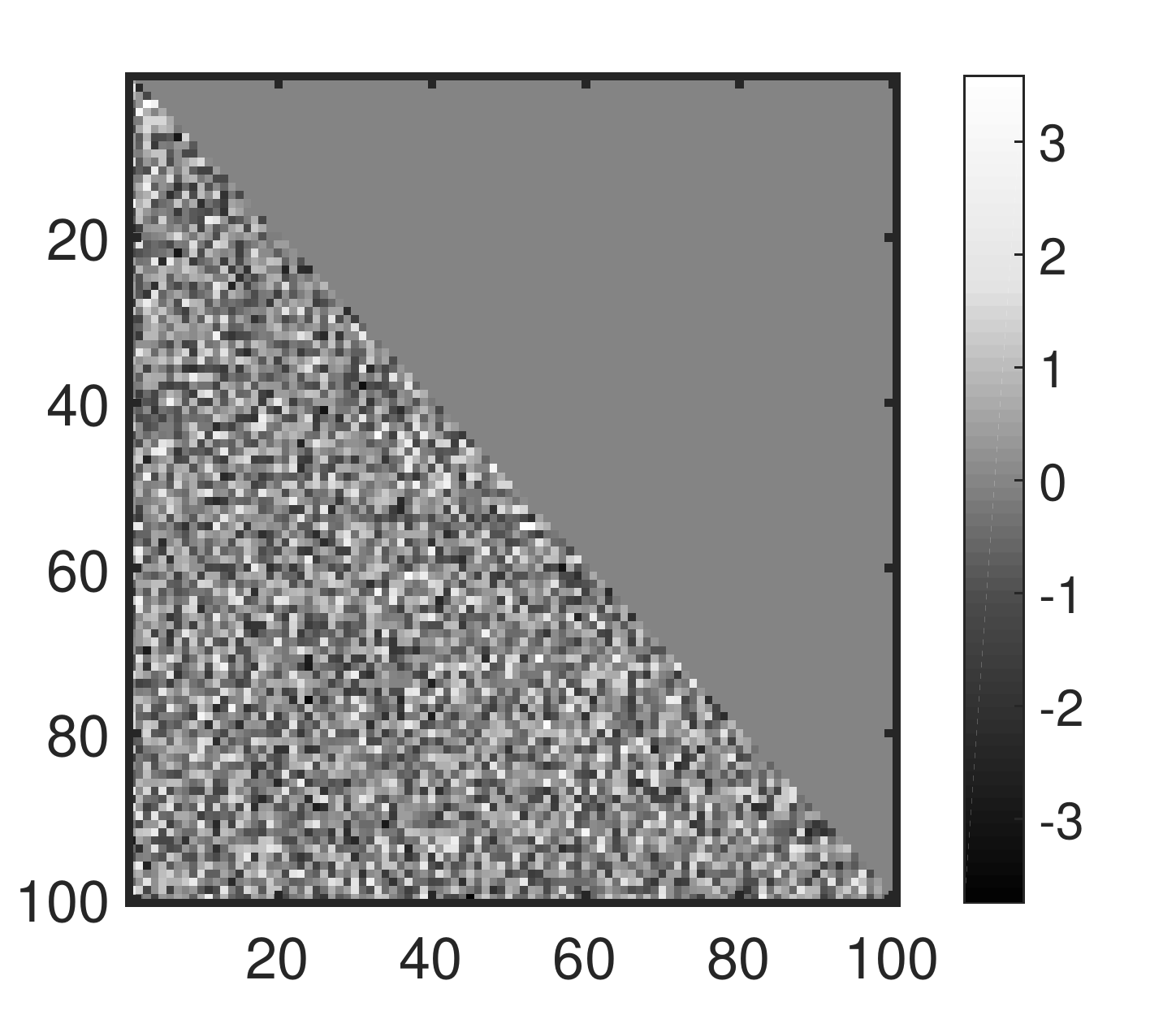}\label{figOneB}}
   \subfigure[ ]{\includegraphics[height=1.25in]{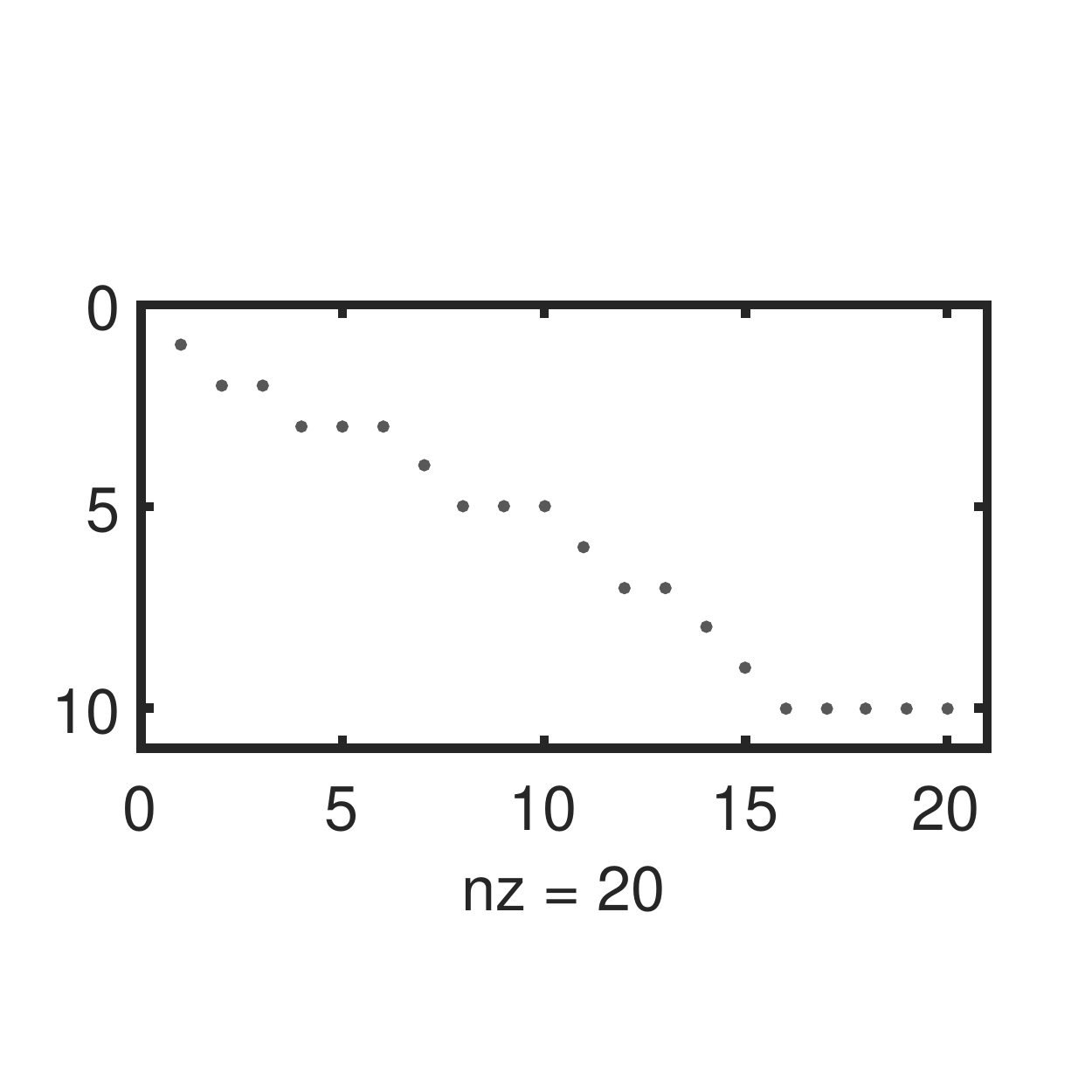}\label{figOneC}}
   \caption{Using age-task frequency matrices to visualize data. This figure shows how the age-task frequency matrix may be used to display data of interest, particularly in the case that there are either a large number of age groups or a large number of tasks under consideration. (a) the non-zero age-task relationships, and (b) the relative values for age-task relationships. The examples shown here are purely illustrative and do not represent age-task frequency matrices for any actual data. A continuous (a), and a discrete (c) caste system in analogy with the figures of E.O.\ Wilson \cite{oster1978}. }
   \label{figOne}
 \end{figure}

One of the more useful aspects of using the age-task frequency matrix to quantify task allocation is the ease with which it allows visualization of the data, particulary in the case of either a large number of age groups, or a large number of tasks. For instance, figure \ref{figOneA} shows the non-zero entries in a matrix with the number of non-zero entries listed. The interpretation being that, each age group performs at-least one specific task, the oldest age group performs every task, although not necessarily with the same frequency, and some age groups perform some number of certain other task which can be ``read off'' of the figure at a glance. If one is interested in comparing relative values for the frequency with which each age group performs a given task, a visualization such as is show in figure \ref{figOneB} would indicate such information. The figures \ref{figOneA} and \ref{figOneC} represent a, respectively, continuous and discrete caste system. These figures may be seen in analogy to the famous figures of Wilson, \emph{cf.} \cite{oster1978}.

In order to determine the dynamics of the frequencies, we establish a coupled system of nonlinear differential equations
\begin{equation}
\frac{d\omega_{ij}}{d \tau} = G_{ij}\cdot\left(\mathcal{L}_{ij} - \mathcal{K}_{ij}\right) , \ i=1,2,\ldots,A;\ j = 1,2,\ldots,T, \label{eq:form} \end{equation}
where $G_{ij}$ is the {\bf frequency growth} function, $\mathcal{L}_{ij}$ is the {\bf likelihood of frequency increase} function, and $\mathcal{K}_{ij}$ is the {\bf likelihood of frequency decrease} function. In general, each of these functions will depend on some subset of the entries of $\Omega$. The function $\mathcal{K}_{ij}$ may be thought of as a penalty term that represents the cost of frequency of performance of task $j$ by age group $i$ for performing tasks other than task $j$.

There are a number of options for choosing the specific form of the frequency growth function $G_{ij}$.  The simplest choice would be for $G_{ij}$ to depend linearly on $\omega_{ij}$. However, we find it convenient, and in somewhat better agreement with observation to take
\begin{equation}
  G_{ij} = r_{ij}\omega_{ij}\left(1 - \sum_{k=1}^{A}\alpha_{ijk}\omega_{kj}\right) , \label{eq:growth}
\end{equation}
where the $r_{ij}$ are rates, and the $\alpha_{ijk}$, $i=1,\ldots,A,$ $j,k=1,\ldots, T$, are weights that specify how sensitive the performance of a given task is to the frequency with which that task is performed by members of all of the age groups. It is necessary to place constraints on the possible values of the weights $\alpha_{ijk}$ to insure that $0\leq \omega_{ij} \leq 1$ is maintained.

The form (\ref{eq:growth}) is arguably the next simplest form to linear dependence, yet which leads to reasonable steady-state conditions. This issue is exemplified below in the setting of a special case. For now, note that the form of $G_{ij}$ implies that the model admits a constant solution of $\omega_{ij} = 0$, for each $i,j$. This is desirable since it is expected that some tasks will not be performed by certain age groups. Furthermore, for some (but not for all possible) values of the weights, $\alpha_{ijk}$, we can expect positive constant solutions. This is an attractive feature since it allows for the possibility that an age-group may carry out a variety of tasks at steady-state frequencies. Thus, when the approach (\ref{eq:form}), with definition (\ref{eq:growth}), is used to model experimentally observed behavior, \emph{a priori} knowledge of expected steady-state behavior should play an important role in putting constraints on the values of the weight parameters. Then, choices for $\mathcal{L}_{ij}$  and $\mathcal{K}_{ij}$ can be made to represent the way that the rate of change of a frequency $\omega_{ij}$ depends on the task needs of the colony, or to admit additional relevant constant solutions. Ultimately, the choice for $G_{ij}$ should be taken based on the specific application and available data.

The likelihood of frequency increase function $\mathcal{L}_{ij}$, and the frequency decrease $\mathcal{K}_{ij}$ will generally both be functions of some of the entries in $\Omega$. Typically, they will each be determined by what is known about the nature of the tasks under consideration, along with hypotheses regarding what steady-state behaviors should be possible.

\section{Model for Two Age Groups and Two Tasks}
\label{firstApplication}

In this section we employ the approach described in section \ref{methods} to model the task allocation dynamics in the case in which there are two distinct tasks which are carried out by two distinct age groups. While this may initially be perceived as an oversimplification of reality, it is actually the case that tasks tend to fall into a small number of broad classifications, such as brood-care or foraging. In fact, this is the most natural case to consider since, to a large extent, this corresponds to the natural division between in-nest and out-of-nest tasks. Thus, one can distinguish a small number of general task classes, and then, if necessary, break these down into a larger number of more specific tasks. Either way, the application given in this section serves to illustrate the implementation of the approach we have developed, and is well worth consideration. In addition, the situation considered in this section is more amenable to direct analysis in ways that a model including either a greater number of tasks, or a greater number of age groups would not be.

Applying equations (\ref{eq:form}) and (\ref{eq:growth}) to the case of two tasks and two age groups leads to a model of the form
\begin{align}
  \frac{d\omega_{11}}{d \tau} = r_{11}\omega_{11}\left(1 - \sum_{k=1}^{2}\alpha_{11k}\omega_{k1}\right)\left(\mathcal{L}_{11} - \mathcal{K}_{11}\right), \label{eq:twoA} \\
  \frac{d\omega_{12}}{d \tau} = r_{12}\omega_{12}\left(1 - \sum_{k=1}^{2}\alpha_{12k}\omega_{k2}\right)\left(\mathcal{L}_{12} - \mathcal{K}_{12}\right), \label{eq:twoB} \\
  \frac{d\omega_{21}}{d \tau} = r_{21}\omega_{21}\left(1 - \sum_{k=1}^{2}\alpha_{21k}\omega_{k1}\right)\left(\mathcal{L}_{21} - \mathcal{K}_{21}\right), \label{eq:twoC} \\
  \frac{d\omega_{22}}{d \tau} = r_{22}\omega_{22}\left(1 - \sum_{k=1}^{2}\alpha_{22k}\omega_{k2}\right)\left(\mathcal{L}_{22} - \mathcal{K}_{22}\right). \label{eq:twoD}
\end{align}
What we need to do now, is to establish appropriate forms for the functions $\mathcal{L}_{ij}$ and $\mathcal{K}_{ij}$. How these may be chosen is developed by way of a sequence of examples in the following subsections.

\subsection{One Specialist: Equilibrium Case}

We begin with a scenario that is as simple as possible, while simultaneously corresponding to task dynamics that can be observed experimentally. This is the case in which one of the age-groups performing the two-tasks is highly specialized and maintains a more or less steady task frequency. More specifically, we assume that age group 1 specializes in task 1, never performs task 2 so that $\omega_{12}\equiv 0$, and furthermore, quickly establishes an equilibrium frequency $\omega_{11}\equiv \bar{\omega}$. Thus, we have two dynamic quantities, $\omega_{22}$, the frequency with which age group 2 performs task 2; and $\omega_{21}$, the frequency with which age group 2 performs task 1. In this scenario, we think of task 2 as being the ``preferred'' task for age group 2. In other words, age group 2 attempts to perform task 2 with maximum frequency, unless there is a significant stimulus, say based on colony need, which causes age group 2 to perform task 1 with some positive frequency. However, age group 2 will also simultaneously try to minimize the frequency with which they perform task 1 unless this is highly detrimental to the overall work requirements of the colony. Examples of two tasks that have features as described above would be if task 1 is brood-care and task 2 is foraging in \emph{Pheidole dentata}.

We note how one could directly derive a model for the situation just outlined. Proceed by assuming that $\omega_{22}$ attempts to increase to a maximum frequency but will simultaneously decrease in proportion to any increase in $\omega_{21}$, although with a rate that depends on the needs of the colony. Furthermore, there is a natural tendency for $\omega_{21}$ to decrease but at a rate that depends on the requirements of the colony. We take the increase of $\omega_{22}$ to be $\rho_{22}\omega_{22}\left(1 -  p_{22}\omega_{22}\right)$ and the decrease to be $q_{22}\omega_{22}\omega_{21}\left(1 - a_{22}\omega_{21} - b_{22}\omega_{11} \right)\left(1 - p_{22}\omega_{22} \right)$. We take the increase for $\omega_{21}$ to be proportional to the decrease in $\omega_{22}$, and the decrease in $\omega_{21}$ to be $q_{21}\omega_{21}\left(1 - g_{21}\omega_{21} - h_{21}\omega_{11} \right)$. This leads to
\begin{align}
  \frac{d\omega_{22}}{d\tau} & = \rho_{22}\omega_{22}\left(1 -  p_{22}\omega_{22}\right) \nonumber \\
                             & - q_{22}\omega_{22}\omega_{21}\left(1 - a_{22}\omega_{21} - b_{22}\omega_{11} \right)\left(1 - p_{22}\omega_{22} \right), \\
            & = \omega_{22}\left(1 -  p_{22}\omega_{22}\right)\left(\rho_{22} - q_{22}\omega_{21}\left(1 - a_{22}\omega_{21} - b_{22}\omega_{11} \right)\right), \label{eq:TTa}\\
  \frac{d\omega_{21}}{d\tau} & = \rho_{21}\omega_{21}\left(1 - g_{21}\omega_{21} - h_{21}\omega_{11} \right)\omega_{22}(1-p_{22}\omega_{22}) \nonumber \\
            & - q_{21}\omega_{21}\left(1 - g_{21}\omega_{21} - h_{21}\omega_{11} \right), \\
            & =\omega_{21}\left(1 - g_{21}\omega_{21} - h_{21}\omega_{11} \right)\left(\rho_{21}\omega_{22}\left(1 -  p_{22}\omega_{22}\right) - q_{21} \right) \label{eq:TTb}.
  \end{align}

The terms in the equations above are chosen as they are based on the following assumptions: if $\omega_{22}$ is zero, there should be neither an increase or a decrease in $\omega_{22}$. Furthermore, if $\omega_{21}$ is zero, then there should be neither a decrease in $\omega_{22}$ or an increase in $\omega_{21}$. In addition, if age group 1 is performing task 1 at a sufficiently high frequency, or the frequency $\omega_{21}$ is sufficiently high, in accordance with colony need, then $\omega_{22}$ should decrease slowly, if at all. These together say that if task 1 is sufficiently represented then the rate at which $\omega_{22}$ decreases and $\omega_{21}$ increases should be small. On the other hand, if task 2 is sufficiently represented, then it is expected that the rate at which $\omega_{22}$ increases or decreases should be small.

\begin{figure}[h!]
  \centering
  \includegraphics[width=2in,height=2in]{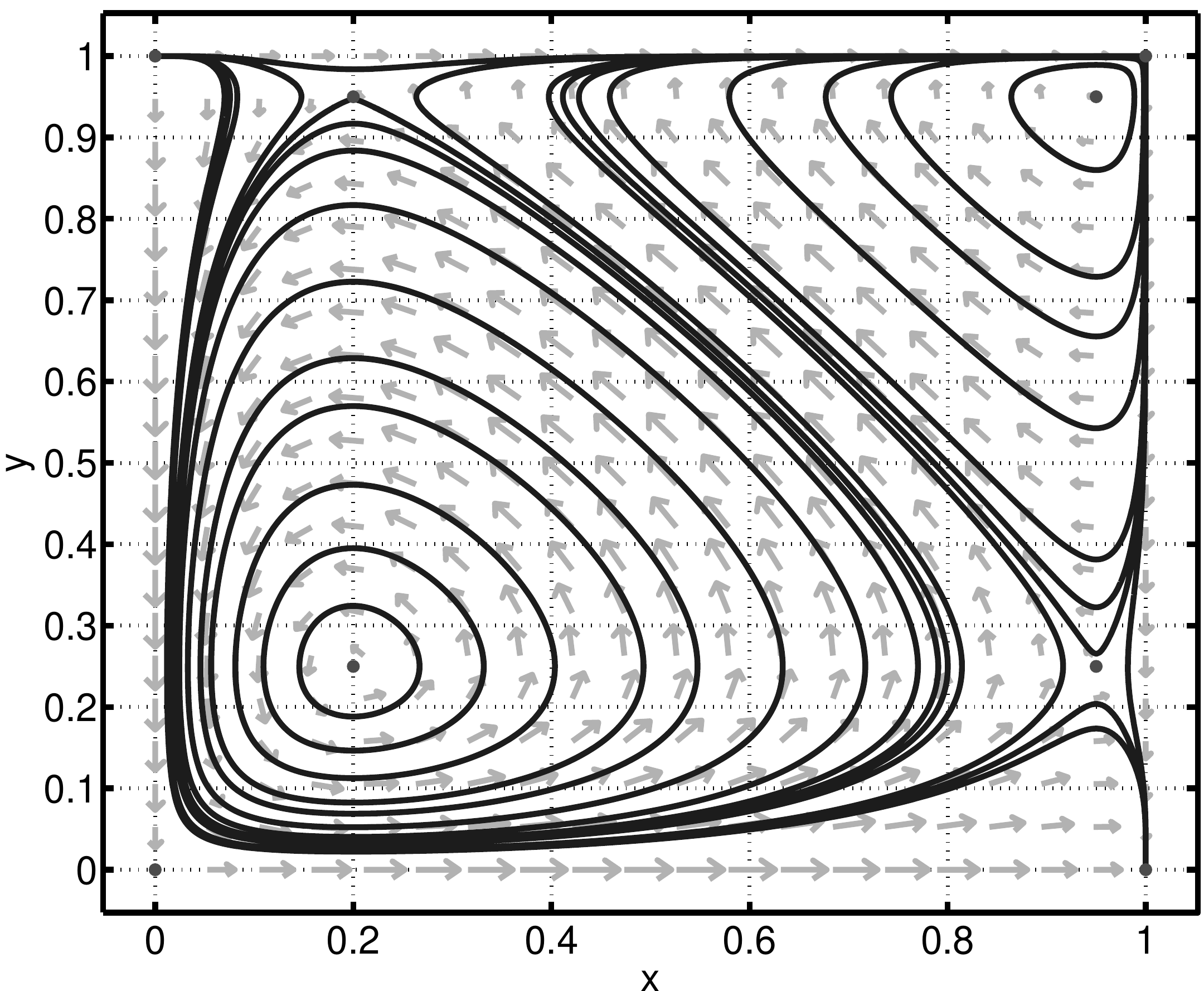}
  \caption{Phase portrait for model (\ref{eq:name8})-(\ref{eq:name9}). This should be compared with the experimental data from \cite{seid2006} which is reproduced in an easily comparable format in figure \ref{oldDynamics}.}\label{modelPhase}
\end{figure}

Finally, observe that the equations (\ref{eq:TTa}), (\ref{eq:TTb}) can be recast in the form of (\ref{eq:twoA})-(\ref{eq:twoD}) in an obvious way, resulting in the two-dimensional system:
\begin{align}
  \frac{d\omega_{22}}{d\tau} & = r_{22}\omega_{22}(1-\alpha_{222}\omega_{22})\left(\mathcal{L}_{22} -\mathcal{K}_{22} \right), \\
                             & = r_{22}\omega_{22}(1-\alpha_{222}\omega_{22})\left(1 - \delta_{22}\omega_{21}(1 - \alpha_{211}\bar{\omega}-\alpha_{212}\omega_{21}) \right),\label{eq:name8} \\
  \frac{d\omega_{21}}{d\tau} & = r_{21}\omega_{21}(1 - \alpha_{211}\bar{\omega}-\alpha_{212}\omega_{21})\left(\mathcal{L}_{21}-\mathcal{K}_{21}  \right), \\
                             & = r_{21}\omega_{21}(1 - \alpha_{211}\bar{\omega}-\alpha_{212}\omega_{21})\left(\gamma_{21}\omega_{22}(1 - a_{222}\omega_{22})  -  \delta_{21} \right).\label{eq:name9}
\end{align}
We seek to explore some of the resulting dynamical possibilities that we believe are relevant to the data from \cite{seid2006}. First, we introduce a simplified notation. Set $x=\omega_{22}$, $y=\omega_{21}$, and $t = \tau$, then the equations (\ref{eq:name8}), (\ref{eq:name9}) can be written as
\begin{align}
  \frac{dx}{dt} & = ax\left(1 - \frac{x}{K_{1}}\right)\left(b - cy\left(L - \frac{y}{K_{2}}\right)\right),\label{eq:simpTTa} \\
  \frac{dy}{dt} & = dy\left(L - \frac{y}{K_{2}}\right)\left(ex\left(1 - \frac{x}{K_{1}}\right) - f\right),\label{eq:simpTTb}
\end{align}
where the parameters in (\ref{eq:simpTTa})-(\ref{eq:simpTTb}) correspond to those in (\ref{eq:name8}) and (\ref{eq:name9}) in the obvious way. Now, observe that, in the system (\ref{eq:simpTTa})-(\ref{eq:simpTTb}), the right hand sides of both equations are each the product of a quadratic function in $x$ times a quadratic function in $y$. Very interesting dynamics can be observed when each of these quadratics admits two real roots in the interval $[0,1]$, where the minimum and maximum root for both $x$ and $y$ is, respectively 0 and 1. In this case, after factoring the quadratic polynomials, our model system takes the form
 \begin{align}
  \frac{dx}{dt} & = Ax(1 - x)(y-y_{1})(y-y_{2}),\label{eq:polyTTa} \\
  \frac{dy}{dt} & = By(1 - y)(x-x_{1})(x-x_{2}),\label{eq:polyTTb}
\end{align}
where the roots $0 < x_{1}\leq x_{2}<1$, $0 <y_{1}\leq y_{2} < 1$, and we allow for $A$ and $B$ to be not necessarily positive numbers. This system then admits up to eight possible steady-state values in the square region $[0,1]\times[0,1]$ of the plane. This is summarized in table \ref{table1}. The appendix includes a more thorough analysis of the system (\ref{eq:simpTTa})-(\ref{eq:simpTTb}). Figure \ref{modelPhase} shows a phase portrait obtained for a model system with this form in the case that there are eight equilibria. We will compare this with experimental results.

The dynamics for (\ref{eq:polyTTa})-(\ref{eq:polyTTb}) shown in figure \ref{modelPhase} are obtained by setting $A=1$, $B=-1$, $x_{1}=\frac{1}{5}$, $x_{2}=\frac{19}{20}$, $y_{1}=\frac{1}{4}$, and $y_{2}=\frac{19}{20}$. We observe that there are three stable steady-states, the asymptotically stable steady-state at $(0,1)$, and the centers at $\left(\frac{1}{5},\frac{1}{4}\right)$ and $\left(\frac{19}{20},\frac{19}{20}\right)$. We interpret the periodic orbits about these two centers as the emergence of repertoire expansion. Note that in there are two distinct patterns with regard to repertoire expansion in this model. Either age-group 2 oscillates around relatively low values for the frequency corresponding to both tasks, or age-group 2 oscillates around relatively high values for the frequency corresponding to both tasks. How does this correspond to the data?

\begin{figure}[h!]
             \centering
             \subfigure{\includegraphics[width=2in,height=2in]{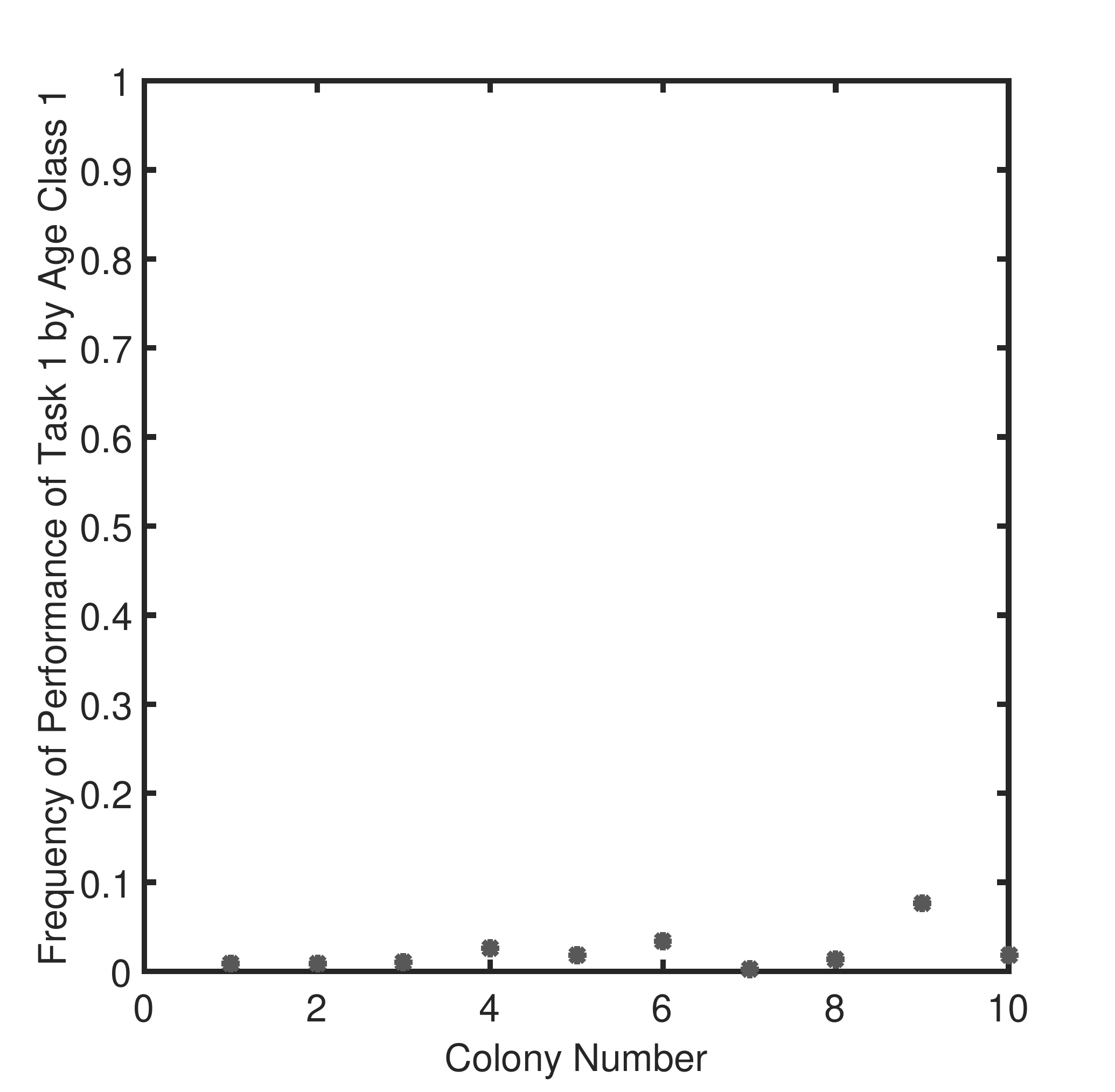}\label{youngSS}}
             \subfigure{\includegraphics[width=2in,height=2in]{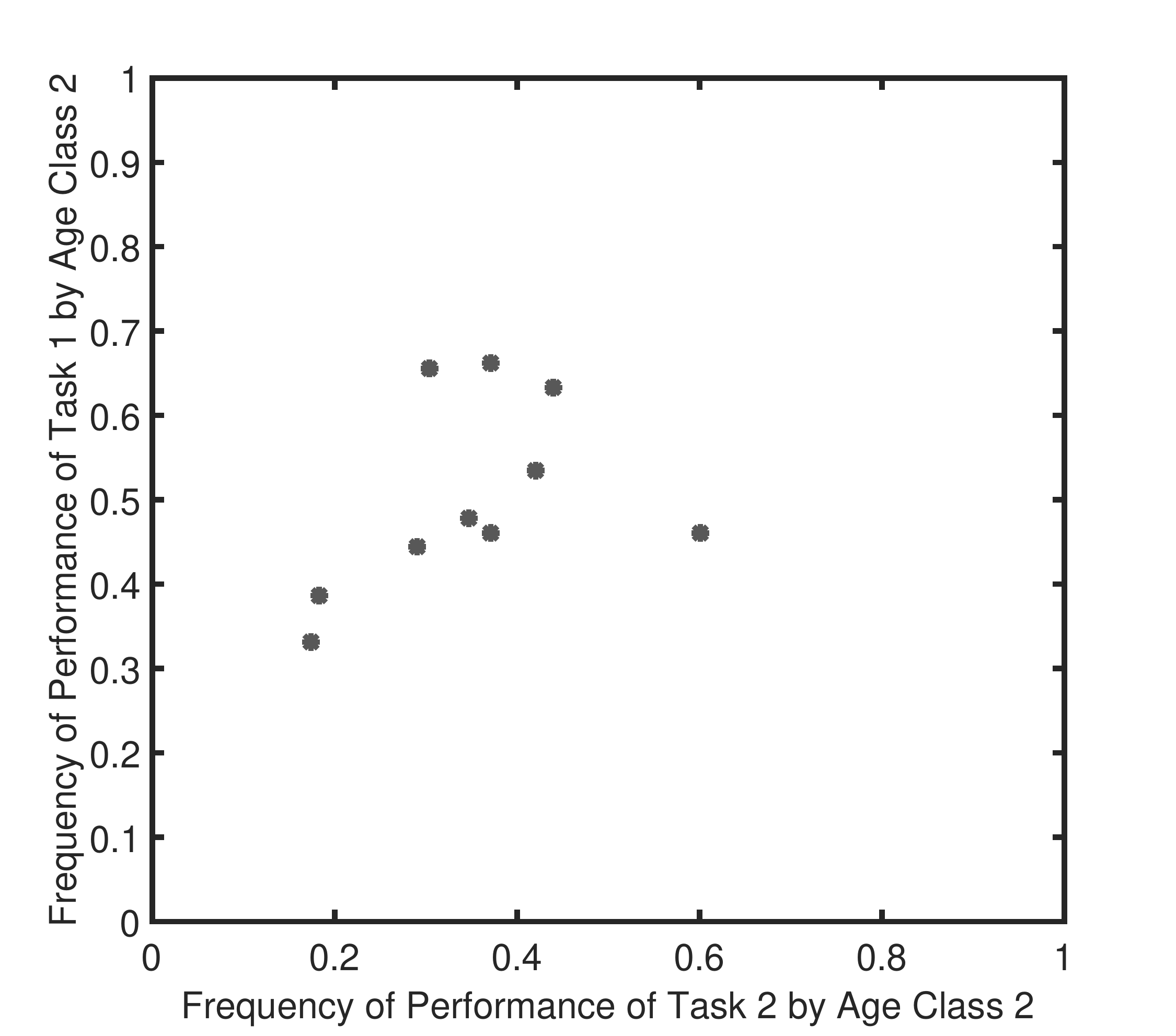}\label{oldDynamics}}
             \caption{Data, from \cite{seid2006}, combined and rescaled to have two tasks and two age groups. The first figure shows the frequency with which the young in each colony carry out their specialized task of ``brood care.'' The second figure shows the plot of the frequency with which old perform their specialized task ``foraging'' versus their secondary task of ``brood care,'' per colony.}\label{phase}
  \end{figure}

In \cite{seid2006}, the authors present data for the relative task performance of four age classes performing nineteen ergonomically distinct tasks with data collected across ten colonies of \emph{Pheidole dentata}. Here, we adapt this data to the case of two distinct age groups, obtained by combining age class A1-A3 of \cite{seid2006} into our age group 1, and taking age-group A4 of \cite{seid2006} as our age-group 2. Furthermore, we combine those tasks from \cite{seid2006} that are well established brood-care acts as our task 1, and those tasks from \cite{seid2006} that are well established foraging acts as our task 2. Figure \ref{phase} shows the frequencies, based on the data from \cite{seid2006}, with which age-group 1 performs their unique task 1 across each of the ten colonies (figure \ref{youngSS}), and a phase diagram for the frequencies of task performances by age-group 2 (figure \ref{oldDynamics}). The phase portrait in figure \ref{phase} should be compared with the results based on experimental data shown in figure \ref{oldDynamics}.

First, note that the data as we have adapted it here exhibits repertoire expansion, or `task stacking'' just as in \cite{seid2006}. Compare figure \ref{expansion} with the figures shown in \cite{seid2006}. Now, the data exhibited in figure \ref{youngSS} suggests that age-group 1 performs only one task, brood-care which corresponds to our task 1, with a relatively low and steady frequency. The data shown in figures \ref{oldDynamics} and \ref{expansion} shows that there is a much more dynamic behavior with respect to the frequency with which age-group 2 performs its tasks. Indeed, the data of figure \ref{oldDynamics} is highly suggestive of the lower left-hand corner of the phase portrait (figure \ref{phase}) corresponding to the model system (\ref{eq:polyTTa})-(\ref{eq:polyTTb}), or equivalently, system (\ref{eq:simpTTa})-(\ref{eq:simpTTb}). Of course, the data from which we have adapted here is an average over time and not truly time-course data. However, one could easily perform experiments to obtain time-course data for a better comparison with the time-dependent model. This is discussed further in the conclusion section below.

 Our model here predicts that, under appropriate circumstances, oscillatory dynamics should be observed if there is repertoire expansion. Furthermore, one should be able to observe two types. Oscillations around two relatively low frequencies of task performance; and oscillations around two relatively high frequencies of task performance.

\subsection{One Specialist: Dynamic Case}
\label{three}

In this section, we extend the results in the previous section by relaxing the assumption that the frequency with which the specialist age-group 1 performs their unique task is in equilibrium. In other words, in the system (\ref{eq:twoA})-(\ref{eq:twoD}) we assume that $\omega_{12}\equiv 0$,  but we do not assume that $\omega_{11}$ necessarily maintains a fixed equilibrium value. Following the reasoning in the previous section, we can derive the following model:
\begin{align}
  \frac{d\omega_{11}}{d\tau} & = \left(a_{11}\omega_{11}  \right)\left(1 - \gamma_{11}\omega_{21} - \delta_{11}\omega_{11} \right),\label{eq:name5} \\
  \frac{d\omega_{22}}{d\tau} & = \left(a_{22}\omega_{22} - b_{22}\omega_{22}\omega_{21}\left(1-\alpha_{22}\omega_{21}-\beta_{22}\omega_{11} \right) \right)\left(1 - \gamma_{22}\omega_{22}  \right),\label{eq:name6} \\
  \frac{d\omega_{21}}{d\tau} & = \left(a_{21}\omega_{22}\omega_{21}(1-\alpha_{21}\omega_{22}) - b_{21}\omega_{21} \right)\left(1 - \gamma_{21}\omega_{21} - \delta_{21}\omega_{11} \right),\label{eq:name7}
\end{align}
where in (\ref{eq:name5}) there is no frequency decrease since there is no other task for age-group 1 to perform. Note that (\ref{eq:name5})-(\ref{eq:name7}) can easily be rewritten in the form of (\ref{eq:form}). Before discussing some properties of solutions to this model, we simply the notation by setting $x=\omega_{22}$, $y=\omega_{21}$, and $z=\omega_{11}$. Furthermore, we simplify the parameter names and write (\ref{eq:name5})-(\ref{eq:name7}) as
\begin{align}
  \frac{dx}{dt} & = ax\left(1 - kx\right)\left(1 - py(1 - my - nz)\right) ,\label{eq:threeX} \\
  \frac{dy}{dt} & = by\left(1 - my - nz\right)\left(qx(1-kx) - d\right) ,\label{eq:threeY} \\
  \frac{dz}{dt} & = cz\left(1 - fz - gy\right) .\label{eq:threeZ}
\end{align}

\begin{figure}[h!]
   \centering
   \subfigure[ ]{\includegraphics[height=1.25in]{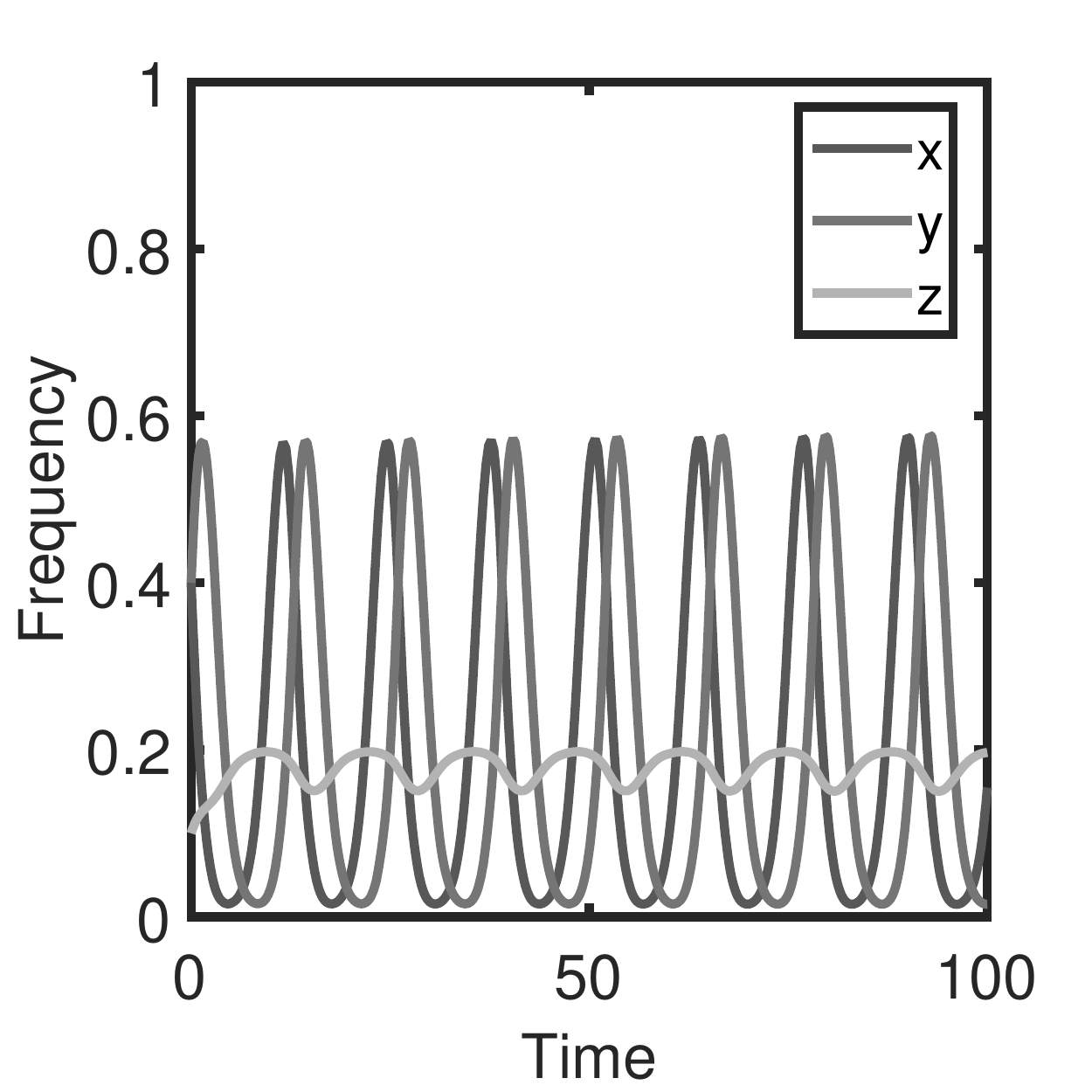}\label{figThreeA}}
   \subfigure[ ]{\includegraphics[height=1.25in]{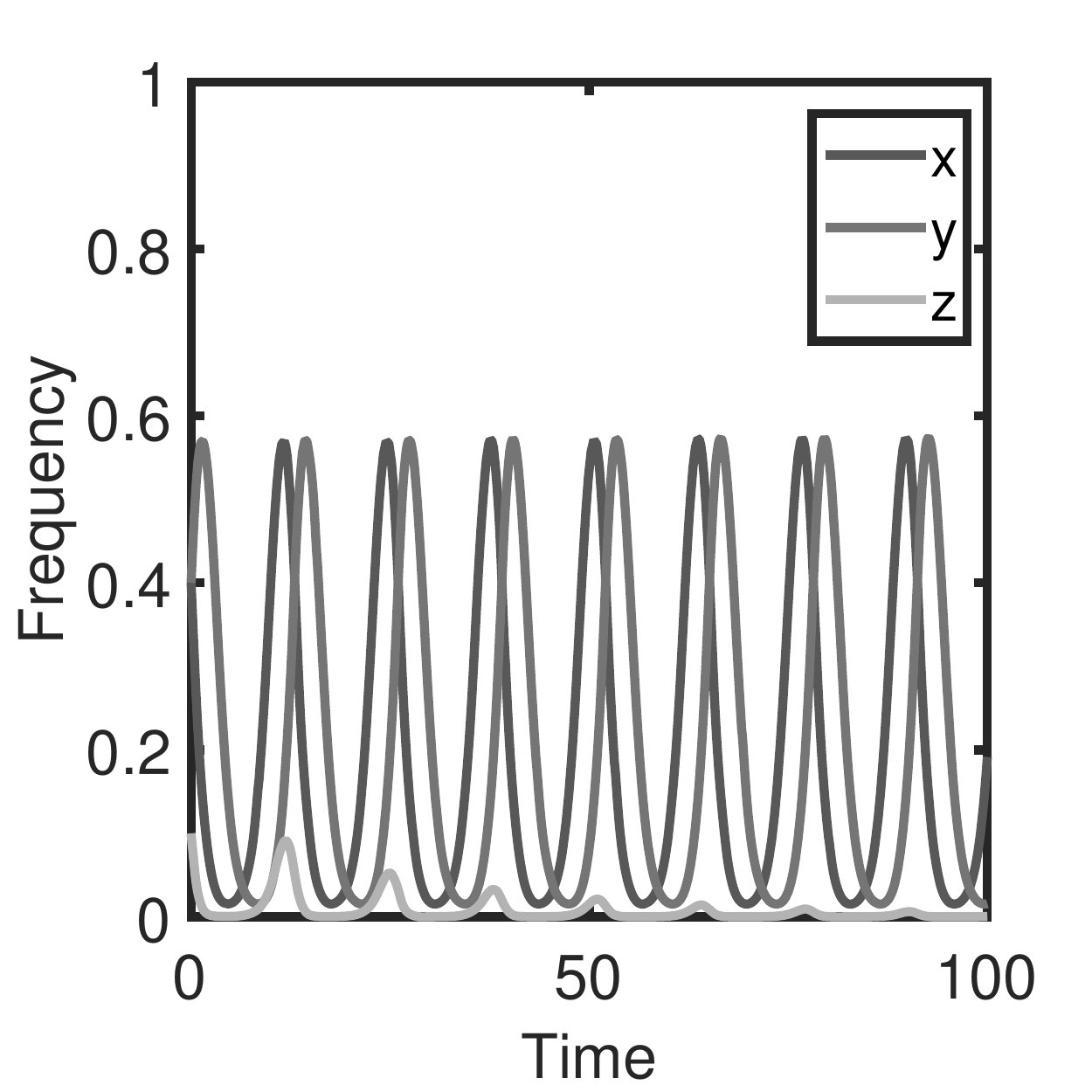}\label{figThreeB}}
   \subfigure[ ]{\includegraphics[height=1.25in]{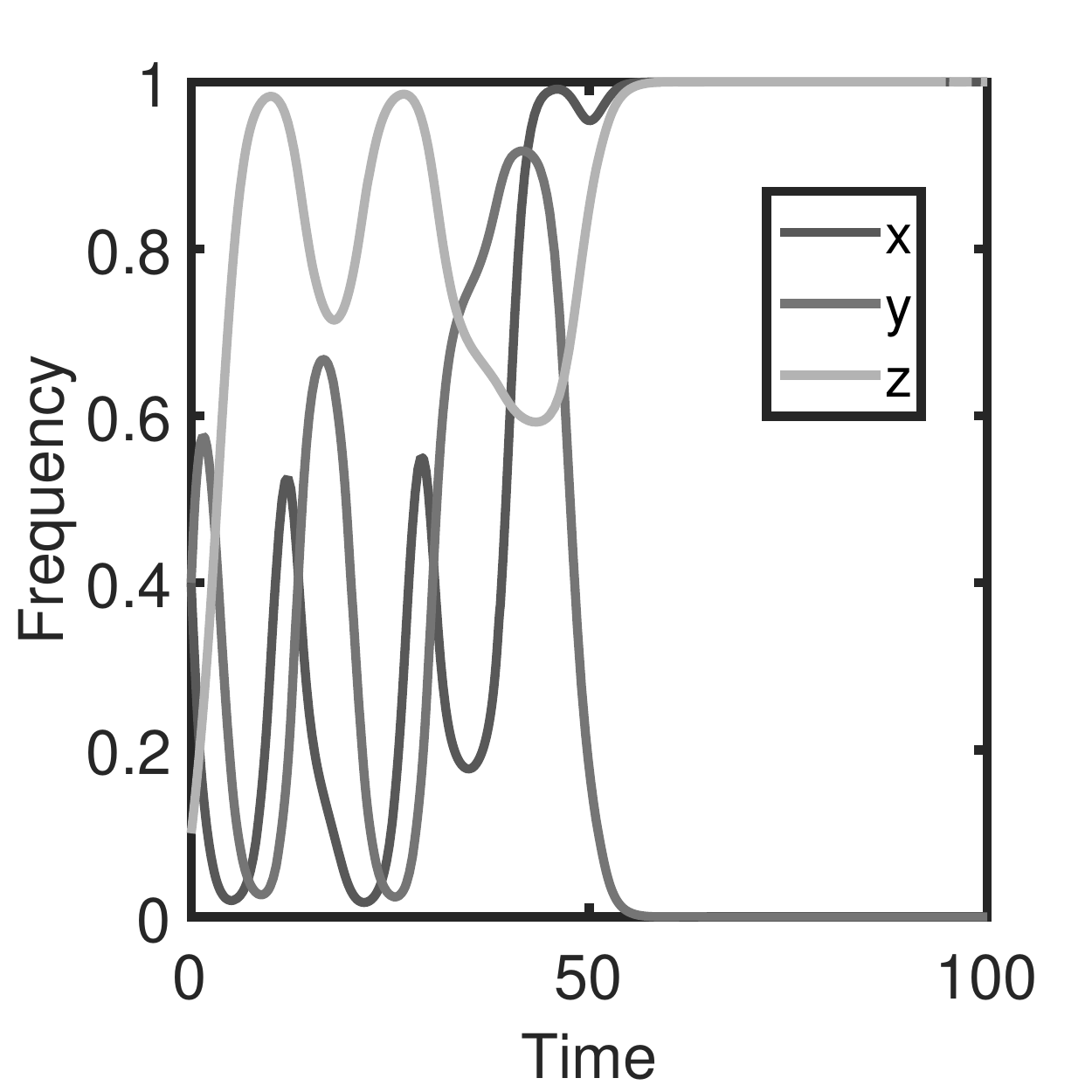}\label{figThreeC}}
   \caption{Some solutions to the example model presented in section \ref{three}. For all of these simulations, the initial conditions for each of the frequencies are kept the same. With the purpose of comparison with experimental results, these initial conditions are taken to be representative of the values shown in figure \ref{figOne}. The three figures are obtained by modifying the weight parameters $m,n,f,g$ from (\ref{eq:threeX})-(\ref{eq:threeZ}). The interpretation of the results are as follows: In figure \ref{figThreeA}, the frequency with which age-group 1 performs its unique task, task 1, oscillates about an equilibrium value. In figure \ref{figThreeB}, the frequency with which age-group 1 performs task 1 initially oscillates but quickly dies out. This is due to the fact that the weight values are such that there is little pressure for age-group 1 to perform. Finally, in figure \ref{figThreeC}, after initial oscillations, the frequency with which age-group 1 performs task 1 and age-group 2 performs task 2 respectively reach a maximum, while the frequency with which age-group 2 performs task 1 dies out. This represents a scenario of transient repertoire expansion, where there is a period in which the needs of the colony are such that age-group 2 must adjust but after some time the requirements are met by age-group 1 performing task 1 and age-group 2 performing only task 2.}
   \label{figThrees}
 \end{figure}

First we observe that (\ref{eq:threeX})-(\ref{eq:threeZ}) will reduce to a model with the form (\ref{eq:name8}), (\ref{eq:name9}) in case $z\equiv 0$. While we do not carry out a full analysis of (\ref{eq:threeX})-(\ref{eq:threeZ}) here, a fuller discussion is presented in the appendix. Note however, that numerical computations will already show that the parameters $m,n,f,g$ play a significant role in determining the dynamics of this system. Recall that the parameters $m,n,f,g$ represent how sensitive the frequencies $y,z$ are to one another. Some results relevant to the phenomena of repertoire expansion are shown in figure \ref{figThrees}.

The three distinct results in figure \ref{figThrees} are obtained by manipulating the weights $m,n,f,g$. Throughout, the initial conditions are kept the same and are set at values representative of the experimental data shown in figure \ref{phase}. In figure \ref{figThreeA}, the frequency with which age-group 1 performs its unique task, task 1, oscillates about an equilibrium value. In figure \ref{figThreeB}, the frequency with which age-group 1 performs task 1 initially oscillates but quickly dies out. This is due to the fact that the weight values are such that there is little pressure for age-group 1 to perform. Finally, in figure \ref{figThreeC}, after initial oscillations, the frequency with which age-group 1 performs task 1 and age-group 2 performs task 2 respectively reach a maximum, while the frequency with which age-group 2 performs task 1 dies out. This represents a scenario of transient repertoire expansion, where there is a period in which the needs of the colony are such that age-group 2 must adjust but after some time the requirements are met by age-group 1 performing task 1 and age-group 2 performing only task 2.

\subsection{General Case}

 Finally, we show how to generalize the models in the previous sections under the assumption that each of the two age groups, the ``young'' (corresponding to $\omega_{11},\omega_{12}$), and the ``old'' (corresponding to $\omega_{21},\omega_{22}$), has a typically preferred task but without assuming an extreme degree of specialization \emph{a priori}. In other words, we suppose that age-group ones specializes in task 1 which would correspond to brood-care, while age-group two specializes in task two corresponding to foraging. Applying reasoning similar to the previous sections, one obtains a model of the form
\begin{align}
  \dot{\omega}_{11}& = \left(a_{11}\omega_{11} - b_{11}\omega_{11}\omega_{12}\left(1-\alpha_{11}\omega_{22}-\beta_{11}\omega_{12} \right) \right)\left(1 - \gamma_{11}\omega_{21} - \delta_{11}\omega_{11} \right),\label{eq:fourA} \\
  \dot{\omega}_{12}& = \left(a_{12}\omega_{11}\omega_{12}(1-\alpha_{12}\omega_{21} - \beta_{12}\omega_{11}) - b_{12}\omega_{12} \right)\left(1 - \gamma_{12}\omega_{22} - \delta_{12}\omega_{12} \right),\label{eq:fourB} \\
  \dot{\omega}_{22}& = \left(a_{22}\omega_{22} - b_{22}\omega_{22}\omega_{21}\left(1-\alpha_{22}\omega_{21}-\beta_{22}\omega_{11} \right) \right)\left(1 - \gamma_{22}\omega_{22} - \delta_{22}\omega_{12} \right),\label{eq:fourC} \\
  \dot{\omega}_{21}& = \left(a_{21}\omega_{22}\omega_{21}(1-\alpha_{21}\omega_{22} - \beta_{21}\omega_{12}) - b_{21}\omega_{21} \right)\left(1 - \gamma_{21}\omega_{21} - \delta_{21}\omega_{11} \right).\label{eq:fourD}
\end{align}

Based on experimental observation, when the tasks are taken as brood-care and foraging, the only realistic solution to (\ref{eq:fourB}) is $\omega_{12} \equiv 0$, which reduces to the case discussed in the previous section. As such, we do not here explore the the complete dynamics of the system (\ref{eq:fourA})-(\ref{eq:fourD}) as this would have no additional bearing to our study of repertoire-expansion.

\section{Conclusion}

We have described a novel approach to the construction of mathematical representations of social insect task allocation via the modeling of the dynamics of the frequencies with which distinct age groups perform distinct tasks. Furthermore, as our main application, we have shown how this model can exhibit the emergence of a particular class of task allocation dynamics known as age-related repertoire expansion. This phenomenon has been experimentally observed, and has been described in the literature, particularly in \cite{seid2008,seid2006}. Additionally, we have shown that, by varying parameters, other classes of task allocation dynamics may result.

The application of our approach presented here is essentially to the task dynamics for brood-care and foraging. Among ants, bees, and other social insects, these tasks are of particular significance since they are fundamental to colony growth and development. However, one may reasonably wish to distinguish between a larger set of more specific tasks. The age-task frequency analysis developed here may be employed in such situations, but there is of course the usual tradeoff between the inclusion of fine detail and simplicity of the mathematical model. Thus there are many directions of elaboration for the techniques presented in this work.

A primary prediction of our model is that, in the case repertoire expansion, one might expect oscillatory dynamics as observed in figure \ref{modelPhase} which is, at least partially, supported by the data from \cite{seid2006}. The analysis of this data is reproduced, albeit in a rescaled and reduced alternative, in figures \ref{expansion} and \ref{phase}. We argue that this type of dynamics fits well with observation and intuition within the scope of repertoire expansion. If task-stacking, such as is exhibited in figure \ref{expansion} or in \cite{seid2006}, is to occur as part of an effective process, or to arise as a regular behavior, stable oscillations seem more probable than does a switch from an unstable source to a stable sink, or some other such dynamics. Greater flexibility and more robust responses to, say environmental, perturbations is provided since there is no requirement of a tendency toward a fixed set of frequencies with which a worker that carries out a variety of different tasks does so. It remains to present experimental and observational evidence in further support of our theoretical ideas.

It is interesting to consider the relation, if any, of the change in frequency of task performance, such as observed in age-related repertoire expansion, with fitness. Here we refer to fitness at the level of the colony since it is at the colony level that natural selection is expected to influence group cooperation among ants. It has been established that colony fitness is related to colony growth rate, see \emph{e.g.}\ \cite{sagili2011}. Thus, it is relevant to seek to determine how the change in frequency of task performance affects colony growth rate. In order to do so in the context of the age-related repertoire expansion data modeled mathematically in this work we require additional observations. However, based on the comparison of our theoretical results with the existing experimental results, we suggest that the correspondence between the dynamics shown in figure \ref{modelPhase} and the experimental result shown in figure \ref{phase}  could be due in large part to the fitness/frequency-change correlation.

To provide additional validation of the modeling approach we have developed, in future work we propose to test our model using two distinct characteristics of \emph{P. dentata}. In \emph{P. dentata}, tasks can be easily grouped and split into two distinct categories, brood-care tasks and non-brood care tasks \cite{seid2005A,seid2005B}, therefore corresponding to the application of our framework that represents such a division of labor. Furthermore, ants are known to readily accept brood (pupa and larva) from conspecific colonies and adopt/care for that brood \cite{holldobler1990}. Thus, we have the ability to manipulate the work-load for a specific group of tasks such as, in the case of our model, brood care. It is our intention to further test our model by manipulating brood number, which should increase the demand for brood care tasks and then measure the effects on task performance of non-brood care tasks. Both our behavioral model and our mathematical model predict that older workers will shift task performance toward brood care at the expense of other tasks. This shift will last until the need for other tasks (\emph{i.e.}\ Foraging) reach a critical level of need or when the brood care tasks needs are met. It will be interesting to compare the empirical result to the modes of oscillations produced by our model.

%
%
%

\section*{Appendix}
\label{analysis}

In this appendix we expand on the analysis of the example models discussed in section \ref{firstApplication}. We begin with a closer examination of the model system (\ref{eq:simpTTa})-(\ref{eq:simpTTb}). What happens in the case that the quadratic function in $y$ in (\ref{eq:simpTTa}), or the quadratic function in $x$ in (\ref{eq:simpTTb}) do not have real roots? In this case it is easy to verify (see the linearization of the system presented below) that the only equilibria are $(0,0),(1,0),(0,1),(1,1)$ and each one will be either a saddle, or a stable or unstable node. No matter, in this case there is only one way that repertoire expansion may arise and that is if (1,1) is a stable node, since otherwise a specific task is selected, and this is not really consistent with any biological observations\footnote{We furthermore rule out the situation in which the equilibria $(0,0),(1,0),(0,1),(1,1)$ are double equilibria for the system.}.

Now, consider again the model from section \ref{firstApplication} in the case that there are eight equilibria. Then, the model can be written as (\ref{eq:polyTTa})-(\ref{eq:polyTTb}). The corresponding Jacobian matrix is then
\begin{equation}
   J_{(x,y)} = \left(\begin{array}{cc} A(1-2x)(y-y_{1})(y-y_{2}) & Ax(1-x)(2y-y_{1}-y_{2}) \\ By(1-y)(2x-x_{1}-x_{2}) & B(1-2y)(x-x_{1})(x-x_{2})  \end{array}\right). \label{eq:jac}
\end{equation}
From this, one can then examine the linearization of the system (\ref{eq:polyTTa})-(\ref{eq:polyTTb}) about each of the equilibria.

We observe the following: For each of the four equilibria $(0,0),(1,0),(0,1),(1,1)$, the matrix (\ref{eq:jac}) becomes
\begin{align}
  J_{(0,0)} &=  \left(\begin{array}{cc} Ay_{1}y_{2} & 0 \\ 0 & Bx_{1}x_{2}  \end{array}\right), \\
  J_{(0,1)} &=  \left(\begin{array}{cc} A(1-y_{1})(1-y_{2}) & 0 \\ 0 & -Bx_{1}x_{2}  \end{array}\right), \\
  J_{(1,0)} &=  \left(\begin{array}{cc} -Ay_{1}y_{2} & 0 \\ 0 & B(1-x_{1})(1-x_{2})  \end{array}\right), \\
  J_{(1,1)} &=  \left(\begin{array}{cc} -A(1-y_{1})(1-y_{2}) & 0 \\ 0 & -B(1-x_{1})(1-x_{2})  \end{array}\right),
\end{align}
from which one sees that
\begin{enumerate}
  \item If $A,B > 0$ then
    \begin{enumerate}
      \item $(0,0)$ is an unstable node;
       \item $(1,0)$ is a saddle;
        \item $(0,1)$ is a saddle;
         \item $(1,1)$ is a stable node.
    \end{enumerate}
  \item If $A > 0, B < 0$ then
    \begin{enumerate}
      \item $(0,0)$ is a saddle;
       \item $(1,0)$ is stable node;
        \item $(0,1)$ is an unstable node;
         \item $(1,1)$ is a saddle.
    \end{enumerate}
  \item If $A < 0, B > 0$ then
    \begin{enumerate}
      \item $(0,0)$ is saddle;
       \item $(1,0)$ is an unstable node;
        \item $(0,1)$ is a stable node;
         \item $(1,1)$ is a saddle.
    \end{enumerate}
  \item If $A,B < 0$ then
    \begin{enumerate}
      \item $(0,0)$ is a stable node;
       \item $(1,0)$ is a saddle;
        \item $(0,1)$ is a saddle;
         \item $(1,1)$ is an unstable node.
    \end{enumerate}
\end{enumerate}

Furthermore, for any of the other four equilibria $(x_{i},y_{j})$, where $i,j\in \{1,2\}$, the Jacobian matrix becomes
\begin{equation}
   J_{(x_{i},y_{j})} = \left(\begin{array}{cc} 0 & Ax_{i}(1-x_{i})(y_{j} - y_{!j}) \\ By_{j}(1-y_{j})(x_{i}-x_{!i}) & 0  \end{array}\right), \label{eq:centerJac}
\end{equation}
where $!i$ and $!j$ denotes the element in the set $\{1,2\}$ that is not equal to the value of $i$ and $j$ respectively. From this, one sees that there may be periodic solutions about an equilibrium $(x_{i},x_{j})$ if the terms $Ax_{i}(1-x_{i})(y_{j} - y_{!j})$ and $By_{j}(1-y_{j})(x_{i}-x_{!i})$ are of opposite sign.

Now we relax the condition of a constant frequency of task performance for age-group 1. Consider again the system (\ref{eq:threeX})-(\ref{eq:threeZ}) from section \ref{three}. Of particular interest is the stability of steady-state solutions to (\ref{eq:threeX})-(\ref{eq:threeZ}) where the equilibrium value for $x$, which we recall represents the frequency of task performance by the specialized age-group 1, is positive and less than one. These are $\left(\frac{1}{k},0,0 \right)$, $\left(\frac{1}{k},0,\frac{1}{f} \right)$, $\left(\frac{1}{k},\frac{1}{m},0 \right)$, $\left(\frac{1}{k},y^{\ast},z^{\ast} \right)$, where $y^{\ast} = \frac{f-nm}{m(f-gn)}$ and $z^{\ast} = \frac{1-g}{f-gn}$. The case that is of most interest in relation to repertoire expansion is the last under the conditions that $k > 1$, $0 < \frac{f-nm}{m(f-gn)} < 1$, and $0 < \frac{1-g}{f-gn} < 1$. In order to facilitate further analysis we consider a rescaling of the equations (\ref{eq:threeX})-(\ref{eq:threeZ}).

Define new variables
\begin{equation}
  \tilde{x} = kx,\ \tilde{y} = my,\ \tilde{z} = nz,\ s = \frac{t}{kbp}. \label{eq:rescale}
\end{equation}
Then under the definition (\ref{eq:rescale}), carrying out the algebra and dropping the tilde's we obtain the rescaled equations
\begin{align}
  \dot{x} & = \alpha x\left(1 - x\right)\left(1 - \rho y(1 - y - z)\right) ,\label{eq:threeXrescale} \\
  \dot{y} & = y\left(1 - y - z\right)\left(x(1-x) - \delta\right) ,\label{eq:threeYrescale} \\
  \dot{z} & = \gamma z\left(1 - \mu z - \nu y\right) ,\label{eq:threeZrescale}
\end{align}
where $\alpha = \frac{a}{kbq}$, $\rho = \frac{p}{m}$, $\delta = \frac{d}{kq}$, $\gamma = \frac{c}{kbq}$, $\mu = \frac{f}{n}$, and $\nu = \frac{g}{m}$; and the equilibrium of principal interest is $(x^{\ast},y^{\ast},z^{\ast}) = \left(1,\frac{\mu - 1}{\mu - \nu},\frac{1 - \nu}{\mu - \nu} \right)$. Then, at $(x^{\ast},y^{\ast},z^{\ast}) = \left(1,\frac{\mu - 1}{\mu - \nu},\frac{1 - \nu}{\mu - \nu} \right)$, we have the linearization
\begin{align}
  J_{\left(x^{\ast},y^{\ast},z^{\ast} \right)} &= \left(\begin{array}{ccc} -\alpha & 0 & 0 \\
  0 & -\delta (1-2 y^{\ast} - z^{\ast}) & \delta y^{\ast} \\
  0 & -\gamma \nu z^{\ast} & \gamma(1 - 2\mu z^{\ast} - \nu y^{\ast})  \end{array} \right),\label{eq:linearOfThreeA} \\
  &=\left(\begin{array}{ccc} -\alpha & 0 & 0 \\
  0 & \delta \frac{\mu-1}{\mu-\nu} & \delta \frac{\mu-1}{\mu-\nu} \\
  0 & -\gamma \nu \frac{1 - \nu}{\mu - \nu}& -\gamma\mu \frac{1-\nu}{\mu-\nu}  \end{array} \right),\label{eq:linearOfThreeB} \\
  &= \left(\begin{array}{ccc} -\alpha & 0 & 0 \\
  0 & \delta y^{\ast} & \delta y^{\ast} \\
  0 & -\gamma \nu z^{\ast} & -\gamma \mu z^{\ast}   \end{array} \right), \label{eq:linearOfThreeC} \\
  & = \left(\begin{array}{ccc} -\alpha & 0 & 0 \\
  0 & A & A \\
  0 & -\nu B & -\mu B   \end{array} \right), \label{eq:linearOfThreeD}
\end{align}
where $A := \delta y^{\ast} =\delta \frac{\mu - 1}{\mu - \nu}$ and $B:=\gamma z^{\ast} = \gamma \frac{1 - \nu}{\mu - \nu}$. First observe that if $(x^{\ast},y^{\ast},z^{\ast})$ is to be a positive steady state then we must have that $A,B > 0$. Now there are two distinct possible cases to consider, when $\mu > \nu$, in which case we must have $\mu > 1$ and $\nu < 1$; and when $\nu > \mu$, in which case we must have $\mu < 1$ and $\nu > 1$. The eigenvalues corresponding to (\ref{eq:linearOfThreeD}) are given by solutions to the cubic $(-\alpha - \lambda)(\lambda^{2} - (A-\mu B)\lambda - (\mu-\nu)AB) = 0$.

Now, the nature of the solutions to the quadratic $\lambda^{2} - (A-\mu B)\lambda - (\mu-\nu)AB = 0$ play a fundamental role in determining the dynamics of solutions to (\ref{eq:threeXrescale})-(\ref{eq:threeZrescale}) about the steady-state $(x^{\ast},y^{\ast},z^{\ast})$ that corresponds to the behavior most relevant to repertoire expansion. In particular, note that if $\mu > \nu$ then there is a possibility for oscillations , and if in addition $\frac{A}{B} = \mu$ there may even be periodic solutions. Since $\mu = \frac{f}{n}$ and $\nu = \frac{g}{m}$ in terms of the original variables from (\ref{eq:threeX})-(\ref{eq:threeZ}) in section \ref{three},  we now see, at least to some extent, the fundamental role that $f,g,m,n$ play in determining the dynamics of this system. Recall that these parameters correspond to how sensitive a given age-task frequency is to the others.

\section*{Acknowledgments}
The authors are grateful to the anonymous referee for the suggestion to include a discussion of the role of fitness as it relates to our work.

\bibliographystyle{plain}
\bibliography{repExpansionRevision}

\newpage

\section{Tables and Figures}

\begin{table}[h!]
   \caption{Summary of possible steady-states and their interpretations in the case of the example model (\ref{eq:polyTTa})-(\ref{eq:polyTTb}).}
   \label{table1}
   \centering
   \begin{tabular}{|c|c|}
     \hline
     Steady-state & Interpretation  \\
     \hline
     $(0,0)$ & Neither task covered by age-group 2 \\
     \hline
     $(1,0)$ & Age-group 2 covers only task 1 \\
     \hline
     $(0,1)$ & Age-group 2 covers only task 2  \\
     \hline
     $(1,1)$ & Age-group 2 performs both tasks at max frequency  \\
     \hline
     $(x_{1},y_{1})$ & Age-group 2 covers both tasks with relatively low frequencies \\
     \hline
     $(x_{1},y_{2})$ & Age-group 2 covers both tasks with task 1 dominant \\
     \hline
     $(x_{2},y_{1})$ & Age-group 2 covers both tasks with task 2 dominant \\
     \hline
     $(x_{2},y_{2})$ & Age-group 2 covers both tasks with relatively high requencies \\
     \hline
   \end{tabular}
 \end{table}

 \newpage

\newpage

 \newpage

\newpage

\newpage

\end{document}